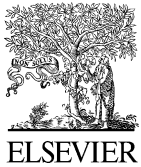

23rd International Symposium on Transportation and Traffic Theory, ISTTT 23, 24-26 July 2019, Lausanne, Switzerland

# Cooperative Adaptive Cruise Control for Connected Autonomous Vehicles by Factoring Communication-Related Constraints

Chaojie Wang[a], Siyuan Gong[b], Anye Zhou[a], Tao Li[c], Srinivas Peeta[a,b,*]

*[a]Civil Engineering, Purdue University, West Lafayette, IN, 47907, USA*
*[b]NEXTRANS, Purdue University, West Lafayette, IN, 47907, USA*
*[c]Computer Science, Purdue University, West Lafayette, IN, 47907, USA*

## Abstract

Emergent cooperative adaptive cruise control (CACC) strategies being proposed in the literature for platoon formation in the Connected Autonomous Vehicle (CAV) context mostly assume idealized fixed information flow topologies (IFTs) for the platoon, implying guaranteed vehicle-to-vehicle (V2V) communications for the IFT assumed. In reality, V2V communications are unreliable due to failures resulting from communication-related constraints such as interference and information congestion. Since CACC strategies entail continuous information broadcasting, communication failures can occur in congested CAV traffic networks, leading to a platoon's IFT varying dynamically. To explicitly factor IFT dynamics and to leverage it to enhance the performance of CACC strategies, this study proposes the idea of dynamically optimizing the IFT for CACC, labeled the CACC-OIFT strategy. Under CACC-OIFT, the vehicles in the platoon cooperatively determine in real-time which vehicles will dynamically deactivate or activate the "send" functionality of their V2V communication devices to generate IFTs that optimize the platoon performance in terms of string stability under the ambient traffic conditions. The CACC-OIFT consists of an IFT optimization model and an adaptive Proportional-Derivative (PD) controller. Given the adaptive PD controller with a two-predecessor-following scheme, and the ambient traffic conditions and the platoon size just before the start of a time period, the IFT optimization model determines the optimal IFT (in terms of the activated and deactivated status of the "send" functionalities of the vehicles in the platoon) that maximizes the expected string stability. This expectation is because each IFT has specific degeneration scenarios whose probabilities are determined by the communication failure probabilities for that time period based on the ambient traffic conditions. The optimal IFT is deployed for that time period, and the adaptive PD controller continuously determines the car-following behaviors of the vehicles based on the unfolding degeneration scenario for each time instant within that period. The effectiveness of the proposed CACC-OIFT is validated through numerical experiments in NS-3 based on NGSIM field data. The results indicate that the proposed CACC-OIFT can significantly enhance the string stability of platoon control in an unreliable V2V communication context, outperforming CACCs with fixed IFTs or with passive adaptive schemes for IFT dynamics.

_______________
 * Corresponding author. Tel: +1-765-496-9726; Fax: +1-765-807-3123.
 *E-mail address:* peeta@purdue.edu

## 1. Introduction

Emerging wireless communication technologies have accelerated the development of connected transportation environments in which vehicles can share information with each other and with surrounding infrastructure. Such information can include safety-related information such as sudden hard braking events, or traffic-related information such as link travel times. Given the wide range of possible applications of vehicle-to-vehicle (V2V) communications using dedicated short-range communication (DSRC) technology, in December 2016, the U.S. National Highway Traffic Safety Administration (NHTSA) proposed to mandate DSRC in new vehicles (NHTSA, 2016). The rapid advances in the autonomous vehicle (AV) industry are also synergistically leading to the leveraging of V2V communication technologies to receive additional information from other connected vehicles and infrastructure. This will provide more opportunities for connected autonomous vehicles (CAVs) to enhance their situational awareness and performance through the implementation of more robust system-level vehicle control strategies, especially platoon-based cooperative adaptive cruise control (CACC) (Nieuwenhuijze et al., 2012).

CACC is an extension of adaptive cruise control (ACC), used to minimize speed differences among vehicles in a platoon and maintain stable and safe headways between adjacent vehicles (Zhou et al., 2017). The CACC literature, discussed hereafter, assumes a platoon with pure (100%) CAVs. Typically, a CACC framework has four components (Li et al., 2015): (i) *node dynamics* (ND), which describes the dynamics of each vehicle in the platoon, such as second-order models (Wang et al., 2014), or third-order models (Guo et al., 2012); (ii) *vehicle-level information flow topology* (VIFT) for a CAV, which describes the configuration of V2V communication links from one CAV to one or more CAVs in the platoon, for example, predecessor-following-leader (Naus et al., 2010) and two-predecessor-following (Zheng et al., 2017) VIFT schemes; (iii) *decentralized controller*, which uses information from other vehicles in the platoon to implement control strategies, such as Proportional-Integral-Derivative (PID) controller (Swaroop et al., 1996), sliding mode controller (Gao et al., 2018), and model predictive controller (Wang et al., 2014; Zhou et al., 2017); and (iv) *formation geometry*, which describes the desired headway between vehicles. Recent studies have modeled the four components of a CAV platoon in different ways, such as the constant distance (CD) policy (Gong et al., 2016), and the constant time headway (CTH) policy (Zhou et al., 2017).

Among the four CACC framework components, the VIFT is closely related to the status of V2V communications. Almost all existing CACC studies assume identical VIFTs for all vehicles in a platoon. As discussed hereafter, consistent with the real world, our study does not constrain the VIFTs to be identical. We label the information flow topology at the platoon level as "IFT," which illustrates the configuration of V2V communication links of all vehicles in the platoon at any time instant. By introducing the time dimension, we consider IFT dynamics, further enhancing modeling realism. By contrast, most studies using a CACC design assume an idealized predetermined, fixed IFT. This assumption ignores the fact that the IFT (and by implication, the VIFTs) can change dynamically due to V2V communication failures (Gao et. al, 2018; Gong et al., 2018). A communication failure may occur due to communication interference or information congestion (Kim et al., 2017; Wang et al., 2017), especially when the ambient (pure CAV) traffic is congested. Information congestion is the reduced quality of service when a communication network node carries more data than it can handle, which in the context of V2V communications is modeled through the potential for failure of information propagation in a V2V communications-enabled traffic network (Wang et al., 2017). Communication interference typically refers to the disruption of a signal as it travels between a sender and a receiver. In the CACC platoon context, though the transmission distance is close enough, interference can arise due to the ambient traffic conditions. A critical reason for interference from other vehicles is the mechanism of the DSRC protocol defined by IEEE 802.11p. In telecommunications, information is transmitted via channels. If more than one sender tries to send information via the same channel at the same time, it will cause interference for both vehicles, and the resulting information collision can cause the transmission to fail. To reduce the probability of information collision, IEEE 802.11p inherits the contention mechanism from IEEE 802.11, which requires every sender to compete for the sending chance. Thereby, the probability of information collision is decreased through this mechanism. However, for V2V communications, if a sender fails to win a sending chance before a new message is generated, the old message will be dropped, and is counted as a communication failure (Qiu et al., 2015). Other factors like the hidden node effect (Jeong et al., 2010) and capture effect (Whitehouse et al., 2005) can also cause communication failures. In all such cases, the distance between some senders is too large for them to sense each other or to keep transmitting signals with enough magnitude so that they can be successfully received. Hence, sender-based communication failure is addressed in this study. By contrast, receiver-related failure is not considered because the spacing between vehicles in a platoon is small enough that a message can be received as long as it is sent successfully.

If communication failures occur, a CACC with a fixed IFT (CACC-FIFT) may execute an erroneous control action or degrade to adaptive cruise control (ACC), which diminishes platoon performance related to mobility, stability, and even safety. To mitigate the negative effects of communication failure, a few studies have proposed novel CACC strategies for a pure CAV platoon by considering dynamic IFT degeneration scenarios. Here, a degeneration scenario for a given IFT refers to any configuration with one or more link communication failures for that IFT. Gong et al. (2018) propose a CACC strategy with dynamic IFT degeneration scenarios (CACC-DIFT), in which a PD controller combined with an acceleration feedforward filter is designed to counteract the IFT dynamics in the platoon. Depending on the IFT degeneration scenario that unfolds at different time instants, the CACC will change the controller parameters to maintain string stability performance rather than degrade to ACC. Gao et al. (2018) design a distributed sliding mode controller based on a linear matrix inequality method to ensure string stability under uncertain but eigenvalue-bounded IFT degeneration scenarios. However, both these studies consider IFT dynamics passively, implying that the controller uses only the functioning links when others have V2V communication failures. While such passive approaches may improve the control performance under unreliable V2V communication networks, their performance is constrained by the ambient traffic conditions which determine the communication failure probabilities of the various IFT links. That is, when the ambient traffic density is higher, the communication failure probabilities are higher. Further, studies (Hafeez et al. 2013) suggest that the IFT in terms of the number of vehicles in the platoon with activated "send" functionality of V2V communication devices within the communication range, is another key factor that impacts communication failures. That is, the communication failure probabilities are higher if several CAVs in the platoon within communication range have their "send" functionality activated. To account for these real-world characteristics, our study proposes the novel idea of proactively controlling in real time the number of platoon vehicles with "send" functionality activated based on the unfolding ambient traffic conditions so as to enhance communication reliability, with the objective of maximizing platoon performance in terms of string stability. We label this strategy the CACC with dynamically optimized IFT, or CACC-OIFT.

Enabling the CACC-OIFT strategy entails addressing some key challenges. First, to account for the time-varying nature of communication failures, all degeneration scenarios of an IFT and their probabilities should be determined. The probability of each degeneration scenario depends on the probability of the communication failure of each link in the IFT, which is itself dynamic and depends on the unfolding traffic conditions. Second, the platoon control performance in terms of string stability needs to be theoretically formulated in the expected sense over all degeneration scenarios for an IFT. While the existing literature uses simulation-based methods (Swaroop et al., 1996; Schakel et al., 2010; Nieuwenhuijze et al., 2012; Zhou et al., 2017) to numerically determine control performance, it is difficult to integrate such approaches in a rigorous optimization model. Third, an adaptive controller is needed to control the car-following behaviors of the vehicles in the CAV platoon based on the unfolding degeneration scenarios for the optimal IFT at different time instants. Hence, there is the need to factor IFT dynamics while ensuring string stability.

The proposed CACC-OIFT strategy for a time period seeks to determine the optimal IFT that maximizes the expected string stability by deactivating or activating the "send" functionality of the V2V communication devices of the vehicles in the platoon, and deploys it for every time instant within that period based on the unfolding degeneration scenarios for that IFT due to V2V communication failures. It includes an IFT optimization model and an adaptive Proportional-Derivative (PD) controller. Fig. 1 illustrates the conceptual flowchart of CACC-OIFT and its operational deployment. Fig. 1(a) shows the various components of CACC-OIFT and their linkages in the time dimension. Given the ambient traffic conditions and platoon size at some time instant $\Delta\tau$ before the start of time period $\tau$ (the period from time instant $t_\tau$ to $t_{\tau+1}$ (i.e. $[t_\tau, t_{\tau+1}]$) in Fig. 1(b)), the IFT optimization model first identifies the candidate IFTs corresponding to the platoon size and their degeneration scenarios. Note that the set of all possible IFTs and their degenerations scenarios is determined offline as they are time invariant and can be predetermined. The subset of IFTs corresponding to the current platoon size denotes the candidate IFT set. Second, the ambient traffic conditions are used to determine the probabilities of the degeneration scenarios for each candidate IFT as these traffic conditions determine the V2V (link) communication failure probabilities. Third, the string stability for each degeneration scenario for each candidate IFT is obtained from the predetermined string stabilities for all degeneration scenarios for all possible IFTs, computed offline using the transfer function in frequency domain of the given adaptive PD controller. This study uses the speed oscillation energy of the platoon as an indicator of string stability performance, which treats the speed oscillation as a signal and computes the sum of speed oscillation energies of all vehicles in the platoon in frequency domain. A lower value for this sum implies that traffic oscillations are damped as they propagate through the platoon, implying better string stability. The optimal IFT (activations and deactivations of "send" functionalities for the platoon vehicles) for time period $\tau$ is determined as the candidate IFT which has the maximum expected string stability across all of its degeneration scenarios.

As shown in Fig. 1(b), the IFT optimization model will determine the optimal IFT for time period $\tau$ at time instant $t_\tau - \Delta\tau$ as it takes $\Delta\tau$ time to solve for the optimal IFT. Note that it is assumed that the ambient traffic conditions and platoon size do not vary within the time period. From Fig. 1(a), the operational deployment of CACC-OIFT starts by proactively deploying the optimal IFT for the first time instant $t_\tau$ of period $\tau$. However, as discussed earlier, due to V2V communication failures, different time instants in period $\tau$ can have a different degeneration scenario of the optimal IFT manifest. The adaptive PD controller continuously determines the car-following behaviors of the vehicles based on the unfolding degeneration scenario for each time instant (i.e. $t_\tau$, $t_\tau + 1$,..., $t_{\tau+1}$) in period $\tau$, thereby controlling vehicular location and dynamics. As shown in Fig. 1(b), at time instant $t_{\tau+1} - \Delta\tau$, the IFT optimization model will update the optimal IFT for the next time period $\tau + 1$. This process continues for the time horizon of interest.

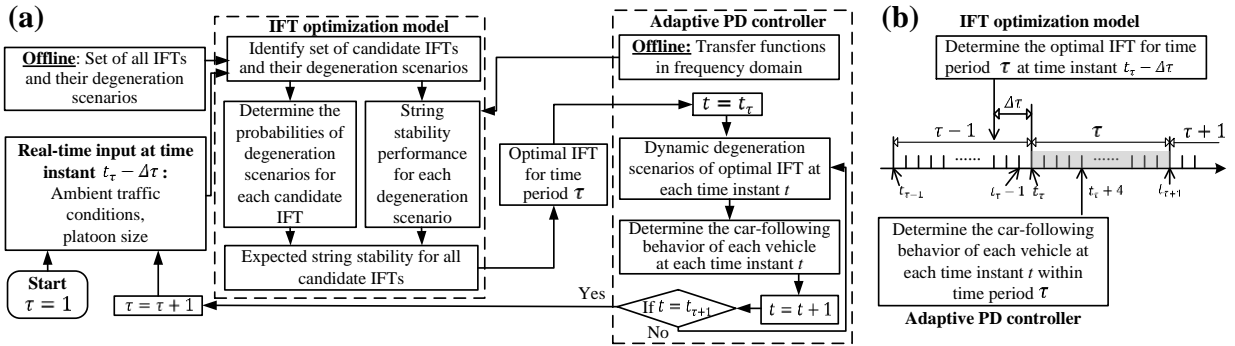

Fig. 1. (a) Conceptual flowchart of CACC-OIFT; (b) Operational deployment of CACC-OIFT.

As the IFT optimization model optimizes the IFT for each time period τ and the adaptive PD controller controls the car-following behaviors for each time instant t, we will formulate the IFT optimization model for one time period in Section 3 and the adaptive PD controller for a time instant within that time period in Section 4. Hence, for notational convenience, we will omit τ and t in these sections. A two-step algorithm is developed to solve the IFT optimization problem, and several critical properties are proved; for example, the leading vehicle in the platoon should always activate its "send" functionality. The effectiveness of the proposed CACC is validated using NGSIM field data (US DOT, 2007) in network simulator NS-3. The results reveal that the algorithm can solve the IFT optimization model for a platoon of considerable size (15 CAVs) in a practically deployable time duration (less than a minute). The proposed CACC-OIFT can significantly damp traffic oscillations and enhance string stability in an unreliable V2V communications context, outperforming CACCs with fixed IFTs or with passive adaptive schemes for IFT dynamics.

The major contributions of the paper are as follows:

(1). We propose an IFT optimization model to explicitly factor IFT dynamics, and leverage it proactively to enhance CACC performance. Compared to passive schemes that simply acknowledge communication failures, a key innovation is to determine the optimal IFT by dynamically and proactively activating or deactivating communication devices of some CAVs in the platoon so as to mitigate negative effects of communication failures and maximize string stability while factoring communication-related constraints.

(2). As another key innovation, the speed oscillation energy in frequency domain is used to evaluate the platoon control performance (i.e. string stability) for a given IFT degeneration scenario. This study treats the speed oscillation as a signal and determines the oscillation energy of each vehicle based on the transfer function of the given controller. The expected oscillation energy for an IFT is the weighted sum of the oscillation energies over all possible degeneration scenarios. Minimizing the expected speed oscillation energy implies maximizing the expected string stability.

(3). To account for the manifestation of different degeneration scenarios of the optimal IFT at different time instants within the time period, we design an adaptive control based on PD feedback controller and acceleration feedforward filter. When the IFT degenerates, the controller determines each vehicle's car-following behavior based on the information it receives at that time instant so that string stability is maintained.

To the best of our knowledge, this is the first study to use rigorous mathematical analysis to improve platoon performance by proactively leveraging IFT dynamics and adjusting adaptive controller parameters. It contributes to the literature in this area and informs the design of CAV platoon control in practice.

The remainder of the paper is organized as follows. Section 2 briefly introduces IFT and degeneration scenarios. Section 3 formulates the IFT optimization model. Section 4 formulates the adaptive controller for an IFT and its degeneration scenarios. Section 5 discusses several critical properties of the proposed CACC-OIFT strategy and discusses the solution algorithm for the IFT optimization problem. Section 6 discusses simulation-based numerical experiments and analyzes the results. Section 7 provides some concluding comments.

## 2. IFT and degeneration scenarios

Though the one-predecessor-following VIFT is the most commonly-used scheme, the more computationally intensive[1] two-predecessor-following VIFT is used in this study to illustrate the CACC-OIFT strategy. Fig. 2 shows a CAV platoon in which a fully-activated two-predecessor-following VIFT is used in the proposed adaptive PD controller. The information of each CAV is delivered to the two vehicles immediately following it through V2V communications. CAV $i$ obtains the state of its two predecessors ($i-1$ and $i-2$), such as location ($x_{i-1}$ and $x_{i-2}$), speed ($\dot{x}_{i-1}$ and $\dot{x}_{i-2}$) and acceleration ($\ddot{x}_{i-1}$ and $\ddot{x}_{i-2}$), through V2V communications. Also, vehicle $i$ can detect the kinematic state ($x_{i-1}$ and $\dot{x}_{i-1}$) of its immediate predecessor $i-1$ through onboard sensors such as radar, Lidar and camera, and its own kinematic state through GPS. The ambient traffic conditions, such as average density $\bar{k}$ and the trajectory oscillation in frequency domain[2] $X(j\omega)$, can be obtained through vehicle-to-infrastructure (V2I) communications, where $\omega$ is the angular frequency, and $j = \sqrt{-1}$.

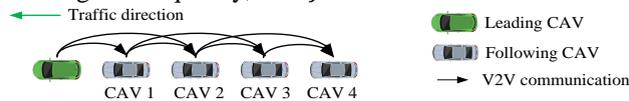

Fig. 2. CAV platoon with a two-predecessor-following VIFT scheme.

Since CACC-OIFT involves dynamically deactivating or activating the "send" functionality of V2V communication devices for vehicles in the platoon, we introduce a vector $\boldsymbol{\xi} = [\eta_0, \eta_1, \dots, \eta_N], \eta_i \in \{0, 1\}$ for $i = 0,1,\dots,N$ to indicate the IFT of a platoon with $N+1$ vehicles, where $\eta_i$ indicates the status of the V2V communication device of vehicle $i$: $\eta_i = 0$, when "send" functionality of V2V communication is deactivated; otherwise, $\eta_i = 1$. For example, the IFT in Fig. 2 has $\boldsymbol{\xi} = [1, 1, 1, 1, 1]$. If some vehicles turn off their "send" functionality, such as vehicles 1, 3 and 4 in Fig. 3(a), the IFT has $\boldsymbol{\xi} = [1, 0, 1, 0, 0]$. We denote $\boldsymbol{\Omega}$ as the set of all possible IFTs that follow the two-predecessor-following scheme.

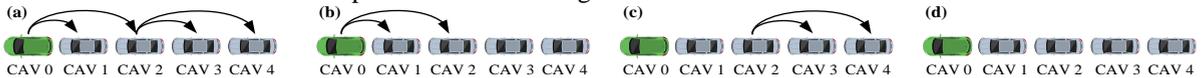

Fig. 3. Example of an IFT and its degeneration scenarios: (a) IFT with "send" functionalities of CAVs 1, 3 and 4 deactivated; (b) Degeneration scenario with CAV 2 failing to send message; (c) Degeneration scenario with CAV 0 failing to send message; (d) Degeneration scenario with both CAVs 0 and 2 failing to send messages.

Though temporarily switching off V2V communications of some vehicles can improve the success rate of other V2V communication links, communication failures cannot be eliminated as they also depend on ambient traffic conditions. As discussed earlier, we focus on failures involving the sending process. Due to sender failure, IFT $\boldsymbol{\xi}$ has degeneration scenarios $\boldsymbol{\xi_d}$ with probabilities $P_d(\boldsymbol{\xi_d})$, which can be formulated in a contention model of V2V communications (Qiu et al., 2015). Here, $d$ is the index of degeneration scenarios, $d = 1, \dots D(\boldsymbol{\xi})$, where $D(\boldsymbol{\xi}) = 2^{\sum_{i=0}^{N} \eta_i}$. The degeneration scenario satisfies $\boldsymbol{\xi_d}(\boldsymbol{\xi}) = [\eta_{0,d}, \eta_{1,d}, \dots, \eta_{N,d}], \eta_{i,d} \in \{0, 1\}, \eta_{i,d} \leq \eta_i$ for $i = 0,1,\dots,N$, which indicates that sender failure only exists for a vehicle with an activated "send" functionality. Hence, degeneration scenario $\boldsymbol{\xi_d}$ is related to IFT $\boldsymbol{\xi}$. For example, the IFT $\boldsymbol{\xi} = [1, 0, 1, 0, 0]$ in Fig. 3(a) has four degeneration scenarios: $\boldsymbol{\xi_1}(\boldsymbol{\xi}) = [1, 0, 1, 0, 0], \boldsymbol{\xi_2}(\boldsymbol{\xi}) = [1, 0, 0, 0, 0], \boldsymbol{\xi_3}(\boldsymbol{\xi}) = [0, 0, 1, 0, 0]$ and $\boldsymbol{\xi_4}(\boldsymbol{\xi}) = [0, 0, 0, 0, 0]$, which are shown in Fig. 3(a)-3(d), respectively. We denote $\boldsymbol{\Omega_d}(\boldsymbol{\xi})$ as the set of all possible degeneration scenarios $\boldsymbol{\xi_d}(\boldsymbol{\xi})$ for IFT $\boldsymbol{\xi}$.

To illustrate the need for controller design, we also analyze the V2V communication status from the receiver side.

---

[1] In a VIFT, if the V2V device sends information to $c$ other devices, it has $2^c$ communication statuses.

[2] The traffic oscillation in frequency domain measures oscillation amplitudes in different frequencies, which can be obtained through Fourier Transform of the ambient vehicles' trajectory data.

Based on different sender failures or deactivated "send" functionalities, a receiver (i.e. vehicle $i$ in Fig. 4) has four possible communication statuses (Fig. 4(a)-4(d)). For a following vehicle $i$, if both predecessors (i.e. $i-1$ and $i-2$) activate communication devices and send information successfully, then the following vehicle $i$ will be controlled by a CACC1 controller with the two-predecessor-following scheme in Fig. 4(a). Fig. 4(b) and 4(c) show the cases where one sender fails to broadcast its message. In these cases, CAV $i$ can detect the kinematic state of its immediate predecessor $i-1$, and obtain one predecessor vehicle's acceleration through V2V communications. When both senders fail (Fig. 4 (d)), CAV $i$ detects the surrounding environment only through onboard sensors. Then, the CACC will degrade to ACC to update the acceleration command based on the relative spacing and speed between CAVs $i$ and $i-1$. Accordingly, a vector is introduced $\zeta(\boldsymbol{\xi_d}) = [\zeta_1(\boldsymbol{\xi_d}(\xi)), \ldots, \zeta_N(\boldsymbol{\xi_d}(\xi))], \zeta_i \in \{1,2,3,4\}$ for $i = 0,1,\ldots,N$ in Equation (1) to indicate receiver status of a platoon with a degeneration scenario $\boldsymbol{\xi_d}$, where $\zeta_i(\boldsymbol{\xi_d}(\xi)) = 1,2,3,4$ indicates that vehicle $i$ is controlled by CACC1, CACC2, CACC3, or ACC, respectively.

$$\zeta(\boldsymbol{\xi_d}(\xi)) = \begin{bmatrix} \zeta_0(\boldsymbol{\xi_d}(\xi)) \\ \zeta_1(\boldsymbol{\xi_d}(\xi)) \\ \vdots \\ \zeta_N(\boldsymbol{\xi_d}(\xi)) \end{bmatrix}^T = \begin{bmatrix} 4 \\ 4 \\ \vdots \\ 4 \end{bmatrix}^T - \boldsymbol{\xi_d}(\xi) \begin{bmatrix} 0 & 2 & 1 & 0 & \cdots & 0 \\ \vdots & 0 & 2 & \ddots & \ddots & \vdots \\ \vdots & \ddots & \ddots & \ddots & \ddots & 0 \\ \vdots & \ddots & \ddots & \ddots & 2 & 1 \\ \vdots & \ddots & \ddots & \ddots & 0 & 2 \\ 0 & \cdots & \cdots & \cdots & \cdots & 0 \end{bmatrix} \qquad (1)$$

Fig. 4. Communication statuses of (a) CACC1; (b) CACC2; (c) CACC3; (d) ACC.

## 3. Formulation of optimization model for the IFT

This section analytically derives the IFT optimization model for a time period. As discussed earlier, due to the dynamics of V2V communication failures, a platoon with IFT $\boldsymbol{\xi}$ may operate under different time-varying degeneration scenarios $\boldsymbol{\xi_d}(\xi)$ at different time instants with corresponding probabilities $P_d(\boldsymbol{\xi_d}(\xi))$. The probabilities of degeneration scenarios of IFT $\boldsymbol{\xi}$ satisfy $\sum_{\boldsymbol{\xi_d}(\xi)\in\Omega_d(\xi)} P_d(\boldsymbol{\xi_d}(\xi)) = 1$. The platoon control performance is a function of the degeneration scenarios. Since degeneration scenario $\boldsymbol{\xi_d}$ is a function of IFT $\boldsymbol{\xi}$, the control performance under it is denoted by $E_d(\boldsymbol{\xi_d}(\xi))$. By considering all possible degeneration scenarios $\Omega_d(\xi)$, the expected platoon control performance $E(\xi)$ for IFT $\boldsymbol{\xi}$ under communication failures is:

$$E(\xi) = \sum_{\boldsymbol{\xi_d}(\xi)\in\Omega_d(\xi)} P_d(\boldsymbol{\xi_d}(\xi)) E_d(\boldsymbol{\xi_d}(\xi)) \qquad (2)$$

Hence, the choice of IFT $\boldsymbol{\xi}$ significantly affects the expected platoon performance, implying the need to determine the IFT that optimizes the expected performance $E'(\xi)$ under the CACC strategy. We summarize the IFT optimization model, denoted as OPT-I, as follows:

$$\begin{aligned} \text{OPT-I} \quad & \underset{\xi\in\Omega}{\text{OPT}}\, E(\xi) = \sum_{\boldsymbol{\xi_d}(\xi)\in\Omega_d(\xi)} P_d(\boldsymbol{\xi_d}(\xi)) E_d(\boldsymbol{\xi_d}(\xi)) \\ \text{s.t.}\quad & \boldsymbol{\xi} = [\eta_0, \eta_1, \ldots, \eta_N], \eta_i \in \{0,1\} \text{ for } i = 0,1,\ldots,N \\ & \boldsymbol{\Omega} = \{[\eta_0, \eta_1, \ldots, \eta_N]|\eta_i \in \{0,1\} \text{ for } i = 0,1,\ldots,N \\ & \boldsymbol{\xi} \in \boldsymbol{\Omega} \\ & \boldsymbol{\xi_d}(\xi) = [\eta_{0,d}, \eta_{1,d}, \ldots, \eta_{N,d}], \eta_{i,d} \in \{0,1\}, \eta_{i,d} \le \eta_i \text{ for } i = 0,1,\ldots,N \\ & \boldsymbol{\Omega_d}(\xi) = \{[\eta_{0,d}, \eta_{1,d}, \ldots, \eta_{N,d}]\, |\, \eta_{i,d} \in \{0,1\}, \eta_{i,d} \le \eta_i \text{ for } i = 0,1,\ldots,N \\ & \sum_{\boldsymbol{\xi_d}(\xi)\in\Omega_d(\xi)} P_d(\boldsymbol{\xi_d}) = 1, \text{ for any } \boldsymbol{\xi} \in \boldsymbol{\Omega} \end{aligned} \qquad (3)$$

The first three constraints of OPT-I relate to the decision variable $\boldsymbol{\xi}$. The first constraint states that $\boldsymbol{\xi}$ is a binary 0-1 vector. The second constraint is the set $\boldsymbol{\Omega}$ of IFTs $\boldsymbol{\xi}$ corresponding to the two-predecessor-following VIFT. The third constraint states that $\boldsymbol{\xi}$ belongs to $\boldsymbol{\Omega}$. The remaining three constraints correspond to the degeneration scenario $\boldsymbol{\xi_d}(\xi)$. The fourth constraint shows the relationship between degeneration scenario $\boldsymbol{\xi_d}(\xi)$ and IFT $\boldsymbol{\xi}$. The fifth constraint indicates that the set $\boldsymbol{\Omega_d}(\xi)$ includes all possible degeneration scenarios for IFT $\boldsymbol{\xi}$. The last constraint states that the probabilities of the degeneration scenarios for an IFT $\boldsymbol{\xi}$ should sum up to 1.

Next, in Section 3.1, the platoon control performance is first characterized in terms of the speed oscillation energy of the vehicles in the platoon. Then, the speed oscillation energy is linked to string stability. Section 3.2 first discusses the determination of the probabilities of the degeneration scenarios. Then, it characterizes the expected string stability for an IFT $\xi$ in terms of the speed oscillation energies of the platoon vehicles. Finally, the IFT optimization model is represented in terms of the expected speed oscillation energies to reflect the expected string stability.

### 3.1 Speed oscillation energy of a degeneration scenario for an IFT and its linkage to string stability

Traffic oscillation is described as "stop and go" or "slow and fast" traffic propagation in traffic flow (Li et al., 2010), which can be measured as a speed oscillation (Li et al., 2012). Traffic oscillation is similar to the case that a signal (i.e. speed oscillation) propagates in a certain medium (i.e. a platoon of vehicles). This study treats speed oscillation as a signal and introduces speed oscillation energy $e_i$ in frequency domain for a vehicle $i$:

$$e_i = \int_0^{+\infty} V_i^2(j\omega)d\omega \tag{4}$$

where $j = \sqrt{-1}$. $V_i(j\omega)$ is the speed oscillation in frequency domain which represents the amplitude of the oscillation with frequency $\omega$. $V_i^2(j\omega)$ is proportional to the energy of this frequency $\omega$. Thereby, the oscillation energy of a vehicle is the sum of its energies in all frequencies. The oscillation energy of all vehicles in a platoon under a degeneration scenario $\xi_d(\xi)$ is given by:

$$E_d\big(\xi_d(\xi)\big) = \sum_{i=0}^N e_i = \sum_{i=0}^N \left[ \int_0^{+\infty} V_i^2(j\omega)d\omega \right] \tag{5}$$

Since a platoon with CACC is an interconnected system, the speed oscillation of each following vehicle will be determined by the speed oscillations of its predecessors and the characteristics of its CACC controller. The relationship between the speed oscillation of the leading vehicle 0 and any following vehicle $i$ can be obtained by recursively linking the speed oscillation of the following vehicle and its predecessors. As discussed in Section 2, the trajectory oscillation information can be directly obtained through V2I communications, unlike speed oscillation information which is an indirect second-order effect. Hence, based on the CACC controller used in degeneration scenario $\xi_d(\xi)$, a transfer function $SS_{X,i}(j\omega,\xi_d(\xi))$ is introduced to measure the propagation of trajectory oscillation in frequency domain from the leading vehicle 0 to any following vehicle $i$:

$$SS_{X,i}\big(j\omega,\xi_d(\xi)\big) = \frac{X_i(j\omega)}{X_0(j\omega)}, \quad i = 1,\dots,n \tag{6}$$

where $X_i(j\omega)$ is the trajectory oscillation of vehicle $i$ in frequency domain. Next, we assume that the trajectory oscillation of the leading vehicle $X_0(j\omega)$ in frequency domain follows the ambient traffic oscillation in frequency domain $X(j\omega)$ $(X_0(j\omega) = X(j\omega))$, since the movement of the leading vehicle is affected by the ambient traffic oscillation.

Given the degeneration scenario $\xi_d(\xi)$, $SS_{X,i}(j\omega,\xi_d(\xi))$ is a function of $\omega$ with a set of predetermined controller parameters. Different transfer functions will be generated under different controllers and IFT degeneration scenarios. Section 4 provide more details on $SS_{X,i}(j\omega,\xi_d(\xi))$.

Next, we derive the speed oscillations in frequency domain, $V_i(j\omega)$, for all vehicles using the leading vehicle trajectory oscillation $X_0(j\omega)$ information. To do so, first, the inverse Fourier transform is performed to obtain the trajectory oscillation of vehicle $i$ in time domain.

$$x_i(t) = \int_0^{+\infty} X_i(j\omega)e^{2\pi t j\omega}d\omega = \int_0^{+\infty} SS_{X,i}\big(j\omega,\xi_d(\xi)\big)X_0(j\omega)e^{2\pi t j\omega}d\omega \tag{7}$$

Then, the derivative of the trajectory oscillation provides the speed oscillation of vehicle $i$ in time domain.

$$v_i(t) = \frac{dx_i(t)}{dt} = \int_0^{+\infty} 2\pi j\omega SS_{X,i}\big(j\omega,\xi_d(\xi)\big)X_0(j\omega)e^{2\pi t j\omega}d\omega \tag{8}$$

Comparing Equation (8) with the inverse Fourier transform of speed oscillation in Equation (9),

$$v_i(t) = \int_0^{+\infty} V_i(j\omega)e^{2\pi t j\omega}d\omega \tag{9}$$

the speed oscillation in frequency domain is as follows:

$$V_i(j\omega) = 2\pi j\omega SS_{X,i}\big(j\omega,\xi_d(\xi)\big)X_0(j\omega) \tag{10}$$

Combining Equations (5) and (10), the speed oscillation energy of the vehicles in the platoon under a degeneration scenario $\xi_d(\xi)$ can be formulated as:

$$E_d\big(\xi_d(\xi)\big) = \sum_{i=0}^{N} V_i^2(j\omega)d\omega = 4\pi^2 \sum_{i=0}^{N} \int_{0}^{+\infty} \omega^2 \mathrm{SS}_{X,i}^2\big(j\omega, \xi_d(\xi)\big) X_0^2(j\omega)d\omega \tag{11}$$

The transfer function $\mathrm{SS}_{X,i}\big(j\omega, \xi_d(\xi)\big)$ is commonly used to infer string stability (Naus et al., 2010). A lower value for $\mathrm{SS}_{X,i}\big(j\omega, \xi_d(\xi)\big)$ reflects better string stability performance as it implies more stability in the traffic flow. From Equation (11), for a given leading vehicle trajectory oscillation $X_0(j\omega)$, a smaller platoon speed oscillation energy $E_d\big(\xi_d(\xi)\big)$ indicates a lower $\mathrm{SS}_{X,i}\big(j\omega, \xi_d(\xi)\big)$ implying better string stability performance.

### 3.2 Probabilities of degeneration scenarios and expected string stability for an IFT

Since the optimal IFT implies that some vehicles have their "send" functionality activated and others have it deactivated, a communication failure implies that an activated "sender" vehicle fails to broadcast its message at a specific time instant. This is a manifestation of a degeneration scenario for this IFT. As discussed in Section 1, sender failure relates to information collision during message broadcasting in the DSRC protocol. This collision occurs when two senders within communication range send information to a receiver at the same time. During the sending process, each sender uses a contention window (CW) to compete for a sending chance. Each sender will randomly select an integer value in the range of $[0, \mathrm{CW}-1]$, to determine when in this window it will send the message. If more than one sender chooses the same integer value, a collision occurs.

In the MAC level protocol (i.e. IEEE 802.11), collision probability can be reduced by increasing the contention window CW size and/or implementing a retransmission scheme. However, both these methods increase time delay, reducing the timeliness of information propagation. For example, if a collision happens, the sender will choose a new contention window and attempts to retransmit the same information after the previous contention window ends (Hafeez et al. 2013). This will lead to delays in the information reaching the receiver, worsening the performance of the real-time controller. Hence, these two methods are not suitable for CACC. Further, as information is generated continuously in the CAV context, if a sender fails to win the sending chance before the next information is generated, the previous information will be dropped, and will therefore also count as a sender failure (Qiu et al. 2015). Therefore, the retransmission scheme is not considered in this study and the contention window CW size will be set to a small value.

In Qiu et al. (2015), a contention model with saturated and unsaturated communication traffic is developed using a Markov chain. The success rate of sending message for vehicle $i$ is:

$$p_{i,\mathrm{sat}} = 2(1 - b_i)(1 - 2b_i + \mathrm{CW})^{-1} \tag{12}$$

where the channel busy rate $b_i$ for sender vehicle $i$ is:

$$b_i = 1 - e^{-\bar{\rho}_i(\xi)p_{i,\mathrm{sat}}} \tag{13}$$

Here, $\bar{\rho}_i(\xi)$ is the average number of vehicles with activated "send" functionalities within communication range $R$ of vehicle $i$. Given the average density $(\bar{k})$ of the ambient traffic flow, the average number of vehicles within the communication range $R$ is given by $m = \lfloor R/\bar{k} \rfloor$. Then, the vector of $\bar{\rho}(\xi) = [\bar{\rho}_0(\xi), \bar{\rho}_1(\xi), \dots, \bar{\rho}_N(\xi)]$ is expressed as:

$$\bar{\rho}(\xi) = \xi M\,(\bar{k}) \tag{14}$$

where, $M\,(\bar{k})$ is a $N$-by-$N$ $2m+1$ diagonal matrix whose non-zero elements are 1 if $m < N$. Otherwise, M is a $N$-by-$N$ matrix in which all elements have the value 1.

Combining Equations (12) and (13), the success rate of sending, $p_{i,\mathrm{sat}}$, can be solved numerically based on a method by Qiu et al. (2012), and for an IFT $\xi$, is denoted as $p_{i,sat}(\xi)$.

Since the V2V device bandwidth and its impact are not considered, a contention model with unsaturated communication is implemented in this study (Qiu et al., 2015). The success rate $p_{i,unsat}$ of a sender vehicle $i$ in one attempt is:

$$p_{i,unsat}(\xi) = [k_1 \log(\bar{\rho}_i(\xi)) + k_2\mathrm{CW} + k_3]p_{i,sat}(\xi) \tag{15}$$

where $k_1, k_2, k_3$ are fitting coefficients.

After obtaining the success rate of a sender vehicle, the probabilities of the degeneration scenarios of an IFT due to sender failure can be calculated. When the IFT degenerates from $\xi$ to $\xi_d$, two sets $A_d(\xi_d(\xi))$ and $B_d(\xi_d(\xi))$ are introduced for the indices of vehicles with successful and unsuccessful broadcast status, respectively, which satisfy $A_d(\xi_d(\xi)) = \{i|\eta_i = 1, \eta_{i,d} = 1, i = 0, \dots, N\}$ and $B_d(\xi_d(\xi)) = \{i|\eta_i = 1, \eta_{i,d} = 0, i = 0, \dots, N\}$. Hence, the probability of the degeneration from $\xi$ to $\xi_d$ is:

$$P_d(\boldsymbol{\xi_d}(\boldsymbol{\xi})) = \prod_{i \in A_d(\boldsymbol{\xi})} p_{i,\text{unsat}} \prod_{i \in B_d(\boldsymbol{\xi})} (1 - p_{i,\text{unsat}}) \qquad (16)$$

The probability of each degeneration scenario is independent of its string stability performance. By substituting Equations (11) and (16) into OPT-I, the optimization model can be reformulated as OPT-II.

OPT-II

$$\min_{\boldsymbol{\xi} \in \boldsymbol{\Omega}} 4\pi^2 \sum_{\boldsymbol{\xi_d}(\boldsymbol{\xi}) \in \boldsymbol{\Omega}_d(\boldsymbol{\xi})} \left[ \prod_{i \in A_d(\boldsymbol{\xi})} p_{i,\text{unsat}} \sum_{i \in B_d(\boldsymbol{\xi})} (1 - p_{i,\text{unsat}}) \sum_{i=1}^{N} \int_0^{+\infty} \omega^2 \text{SS}_{X,i}^2(j\omega, \boldsymbol{\xi_d}(\boldsymbol{\xi})) X_0^2(j\omega) d\omega \right] \qquad (17)$$

s.t. $\quad \boldsymbol{\xi} = [\eta_0, \eta_1, \dots, \eta_N], \eta_i \in \{0, 1\}$ for $i = 0, 1, \dots, N$

$\quad \boldsymbol{\Omega} = \{[\eta_0, \eta_1, \dots, \eta_N] | \eta_i \in \{0, 1\}$ for $i = 0, 1, \dots, N\}$

$\quad \boldsymbol{\xi} \in \boldsymbol{\Omega}$

$\quad \boldsymbol{\xi_d}(\boldsymbol{\xi}) = [\eta_{0,d}, \eta_{1,d}, \dots, \eta_{N,d}], \eta_{i,d} \in \{0, 1\}, \eta_{i,d} \leq \eta_i$ for $i = 0, 1, \dots, N$

$\quad \boldsymbol{\Omega}_d(\boldsymbol{\xi}) = \{[\eta_{0,d}, \eta_{1,d}, \dots, \eta_{N,d}] \mid \eta_{i,d} \in \{0, 1\}, \eta_{i,d} \leq \eta_i$ for $i = 0, 1, \dots, N\}$

$\quad \sum_{\boldsymbol{\xi_d}(\boldsymbol{\xi}) \in \boldsymbol{\Omega}_d(\boldsymbol{\xi})} P_d(\boldsymbol{\xi_d}) = 1, \quad$ for any $\boldsymbol{\xi} \in \boldsymbol{\Omega}$

$\quad A_d(\boldsymbol{\xi_d}(\boldsymbol{\xi})) = \{i | \eta_i = 1, \eta_{i,d} = 1, i = 0, \dots, N\}$

$\quad B_d(\boldsymbol{\xi_d}(\boldsymbol{\xi})) = \{i | \eta_i = 1, \eta_{i,d} = 0, i = 0, \dots, N\}$

OPT-II is a mixed binary-integer optimization problem. It is almost identical to OPT-I, but has two additional constraints at the end related to the successful and unsuccessful broadcast status under the degeneration scenario $\boldsymbol{\xi_d}(\boldsymbol{\xi})$ for IFT $\boldsymbol{\xi}$, respectively. The objective function seeks the IFT whose most likely degeneration scenarios have lower speed oscillation energies (or, higher string stability performance). Section 5 proposes an algorithm to solve OPT-II and discusses its characteristics.

## 4. Formulation of adaptive controller for degeneration scenarios of IFT

We propose an adaptive Proportional-Derivative (PD) controller based on a two-predecessor-following scheme. First, we introduce the control structure (including vehicle dynamics, spacing policy, and feedforward and feedback sub-controllers) that can adapt to any IFT and its degeneration scenarios. Next, we determine several critical parameters to ensure the head-to-tail string stability of the platoon and improve the capability for measurement noise mitigation for individual vehicles. The transfer function $\text{SS}_{X,i}(j\omega, \boldsymbol{\xi_d}(\boldsymbol{\xi}))$ is used in the IFT optimization model as the indicator of string stability performance for all degeneration scenarios $\boldsymbol{\xi_d}(\boldsymbol{\xi})$ of IFT $\boldsymbol{\xi}$.

*4.1 Control structure*

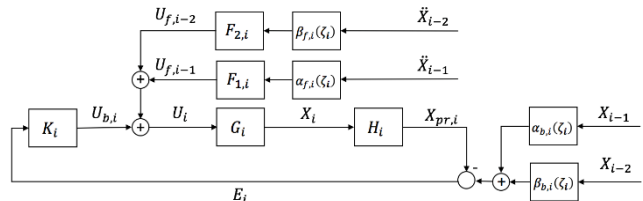

Fig. 5. Schematic of the adaptive PD controller.

The schematic of the adaptive PD controller for vehicle $i$ is illustrated in Fig. 5. $U_i$ represents the control command, which consists of control feedback $U_{b,i}$ from the spacing error $E_i$ and two extra feedforward terms $U_{f,i-1}$ and $U_{f,i-2}$ from the acceleration rates $\ddot{X}_{i-1}$ and $\ddot{X}_{i-2}$, respectively. In the case of ACC, $U_i$ merely consists of a feedback control command. $X_i$ is the position output of vehicle $i$, $X_{pr,i}$ is the processed position output of vehicle $i$ after considering the spacing policy, $X_{i-1}$ is the feedback position information from the immediate predecessor while $X_{i-2}$ is the feedback position information from the second predecessor. $K_i$ is the feedback controller which generates a control command to rectify the spacing error. $G_i$ represents the ideal longitudinal vehicle dynamics. $H_i$ denotes the spacing policy (i.e. such as CD or CTH), and $F_{1,i}$ and $F_{2,i}$ are feedforward filters to process the acceleration information from the corresponding predecessors. $\alpha_{b,i}(\zeta_i)$ and $\beta_{b,i}(\zeta_i)$ are weighting coefficients for position feedback information while $\alpha_{f,i}(\zeta_i)$ and $\beta_{f,i}(\zeta_i)$ are weighting coefficients for acceleration feedforward information. These coefficients are determined by the dynamic IFT that satisfies Equations (18)-(21). Recall that $\zeta_i$ represents the controller status in Equation (1), and $\zeta_i = 1,2,3,4$ indicates that vehicle $i$ is controlled by CACC1, CACC2, CACC3 and ACC, respectively.

$$\alpha_{f,i}(\zeta_i) = \alpha_{b,i}(\zeta_i) = \alpha, \ \beta_{f,i}(\zeta_i) = \beta_{b,i}(\zeta_i) = \beta, \ \alpha + \beta = 1, \ 0 < \alpha, \beta < 1, \quad \text{if } \zeta_i = 1 \text{ for CACC1} \quad (18)$$

$$\alpha_{f,i}(\zeta_i) = \alpha_{b,i}(\zeta_i) = 1, \ \beta_{f,i}(\zeta_i) = \beta_{b,i}(\zeta_i) = 0, \quad \text{if } \zeta_i = 2 \text{ for CACC2} \quad (19)$$

$$\alpha_{f,i}(\zeta_i) = \beta_{b,i}(\zeta_i) = 0, \ \beta_{f,i}(\zeta_i) = \alpha_{b,i}(\zeta_i) = 1, \quad \text{if } \zeta_i = 3 \text{ for CACC3} \quad (20)$$

$$\alpha_{f,i}(\zeta_i) = \beta_{f,i}(\zeta_i) = \beta_{b,i}(\zeta_i) = 0, \ \alpha_{b,i}(\zeta_i) = 1, \quad \text{if } \zeta_i = 4 \text{ for ACC} \quad (21)$$

### 4.1.1. Vehicle dynamics

This study considers an ideal vehicle dynamics model, ignoring air drag, rolling resistance and actuator delay (Li et al., 2015) in the vehicle dynamics model. The linearized state-space representation of the idealized longitudinal vehicle dynamics can be represented as:

$$\dot{x}_i(t) = v_i(t), \qquad \dot{v}_i(t) = u_i(t) \quad (22)$$

where $x_i(t)$, $v_i(t)$, and $u_i(t)$ are the absolute position, velocity and acceleration of vehicle $i$ at time $t$, respectively.

To analyze stability performance, the modeling and analysis are performed in the Laplace domain. The idealized longitudinal vehicle dynamics model in the Laplace domain can be described by using a transfer function:

$$G_i(s) = X_i(s)U_i(s)^{-1} = s^{-2} \quad (23)$$

where the input $U_i(s)$ denotes the acceleration of vehicle $i$ and the output $X_i(s)$ denotes the absolute position of vehicle $i$ in the Laplace domain.

### 4.1.2. Spacing policy

To achieve more efficient damping oscillation, we obtain the desired relative distances between vehicle $i$ and its two predecessors using the CTH policy as follows:

$$d_{i,1}(t) = L + h\dot{x}_i(t), \qquad d_{i,2}(t) = 2[L + h\dot{x}_i(t)] \quad (24)$$

where $d_{1,i}(t)$ is the desired relative distance between vehicle $i$ and vehicle $i-1$, and $d_{2,i}(t)$ is the desired relative distance between vehicle $i$ and vehicle $i-2$. $L$ is the constant standstill distance (including vehicle length) between the two vehicles, $\dot{x}_i(t)$ is the velocity of vehicle $i$, and $h$ is the desired time headway.

The weighted sum of spacing error in Equation (25) is implemented in the feedback loop since the feedback controller processes spacing errors from both predecessors.

$$e_i(t) = \alpha_{b,i}(\zeta_i)\{x_{i-1}(t) - x_{pr,i,1}(t)\} + \beta_{b,i}(\zeta_i)\{x_{i-2}(t) - x_{pr,i,2}(t)\} \quad (25)$$
$$= \alpha_{b,i}(\zeta_i)\{[x_{i-1}(t) - x_i(t)] - d_{i,1}(t)\} + \beta_{b,i}(\zeta_i)\{[x_{i-2}(t) - x_i(t)] - d_{i,2}(t)\}$$

From Equations (18)-(21), we note that $\alpha_{b,i}(\zeta_i) + \beta_{b,i}(\zeta_i) = 1$ is always satisfied in the four controller statuses. Substituting Equation (24) into Equation (25), the weighted spacing error is

$$e_i(t) = \alpha_{b,i}(\zeta_i)x_{i-1}(t) + \beta_{b,i}(\zeta_i)x_{i-2}(t) - x_i(t) - (2 - \alpha_{b,i}(\zeta_i))(L + h\dot{x}_i(t)) \quad (26)$$

Taking the Laplace transformation of Equation (26), the spacing error can be expressed equivalently as:

$$E_i(s) = \alpha_{b,i}(\zeta_i)X_{i-1}(s) + \beta_{b,i}(\zeta_i)X_{i-2}(s) - H_i(s)X_i(s) \quad (27)$$

where $H_i(s)$ is the CTH spacing policy in frequency domain, given by:

$$H_i(s) = 1 + \left(2 - \alpha_{b,i}(\zeta_i)\right)hs \quad (28)$$

### 4.1.3. Acceleration feedforward

From Fig. 5, the relationship between tracking error $E_i(s)$ and the feedforward acceleration $\ddot{X}_{i-1}(s) = s^2X_{i-1}(s)$ of the predecessor $i-1$ and feedforward acceleration $\ddot{X}_{i-2}(s) = s^2X_{i-2}(s)$ of the predecessor $i-2$ in the Laplace domain is formulated as:

$$E_i(s) = \frac{\alpha_{b,i}(\zeta_i) - \alpha_{f,i}(\zeta_i)H_i(s)G_i(s)F_{1,i}(s)s^2}{1 + H_i(s)G_i(s)K_i(s)}X_{i-1}(s) + \frac{\beta_{b,i}(\zeta_i) - \beta_{f,i}(\zeta_i)H_i(s)G_i(s)F_{2,i}(s)s^2}{1 + H_i(s)G_i(s)K_i(s)}X_{i-2}(s) \quad (29)$$

To eliminate spacing error between adjacent vehicles, feedforward filters $F_{1,i}(s)$ and $F_{2,i}(s)$ are designed based on a zero-error condition (Naus et al., 2010). Hence, the numerators of the first and second terms on the right hand side in Equation (29) should be zero. By combining Equations (18) and (23), and the spacing policy $H_i(s)$ in Equation (28), we can derive the feedforward filters as:

$$F_{1,i}(s) = F_{2,i}(s) = (H_i(s)G_i(s)s^2)^{-1} = \left(H_i(s)\right)^{-1} \quad (30)$$

### 4.1.4. Control command

As illustrated in Fig. 5, our control command consists of a feedback term and two feedforward terms:

$$U_i(s) = U_{b,i}(s) + U_{f,i-1}(s) + U_{f,i-2}(s) \quad (31)$$

Recall that the feedback term $U_{b,i}(s)$ uses spacing error to stabilize the closed-loop system while the feedforward terms $U_{f,i-1}(s)$ and $U_{f,i-2}(s)$ use acceleration rates from the two predecessors to minimize the spacing error.

The feedback term $U_{b,i}(s)$ and the corresponding PD feedback controller are defined as:

$$U_{b,i}(s) = K_i(s)E_i(s) \tag{32}$$

$$K_i(s) = \omega_{K,i}(\omega_{K,i} + s) \tag{33}$$

where $\omega_{K,i}$ is the cut-off frequency of the PD controller. Since $\omega_{K,i}$ affects the value of string stability $SS_{X,i}(j\omega, \xi_d(\xi))$ in Equation (6), it has a strong impact on the string stability of the platoon, and will be determined analytically in Section 4.2. $E_i(s)$ is the spacing error in the Laplace domain in Equation (27).

The feedforward terms $U_{f,i-1}(s)$ and $U_{f,i-2}(s)$ indicate that the acceleration rates of vehicles $i-1$ and $i-2$ are sent to $i$, respectively:

$$U_{f,i-1}(s) = \alpha_{f,i}(\zeta_i)F_{1,i}(s)s^2 X_{i-1}(s) \tag{34}$$

$$U_{f,i-2}(s) = \beta_{f,i}(\zeta_i)F_{2,i}(s)s^2 X_{i-2}(s) \tag{35}$$

Note that according to the two-predecessor-following scheme, the second vehicle in the platoon can receive acceleration information from only the leading vehicle, that is, the feedforward term of vehicle 1 includes only $U_{f,0}(s)$.

The overall control command is obtained by summing up Equations (32), (34) and (35). Through inverse Laplace transformation, the expression for the control command is:

$$u_i(t) = \omega_{K,i}^2 e_i(t) + \omega_{K,i}\dot{e}_i(t) + \alpha_{f,i}(\zeta_i)F_{1,i}(t)\ddot{x}_{i-1}(t) + \beta_{f,i}(\zeta_i)F_{2,i}(t)\ddot{x}_{i-2}(t) \tag{36}$$

The discretized version of Equation (36) is used for operational deployment. We do not discuss it here due to space constraints.

## 4.2. Stability analysis and parameter determination

Two parameters in the designed system significantly impact the platoon performance: the time headway $h$ and cut-off frequency $\omega_{K,i}$. This section analyzes the inputs for these parameters to mitigate measurement noise while ensuring the string stability of the platoon.

### 4.2.1. Measurement noise mitigation

Measurement noise is usually a high frequency noise generated from onboard sensors that produces inaccurate trajectory information, leading to undesirable control inputs. Hence, the mitigation of measurement noise effect is essential to improve control performance in terms of stability for individual vehicles in the platoon. Lemma 1 presents the characteristics of an upper bound for the product of $h$ and $\omega_{K,i}$ (i.e. $h\omega_{K,i}$) to mitigate measurement noise.

**Lemma 1:** By setting an upper bound for the product of $h$ and $\omega_{K,i}$ as: $h\omega_{K,i} \leq W_{max}$, the high frequency measurement noise from the two predecessors is individually attenuated by at least $W_{max}/[1 + W_{max}]$.

**Proof:** For any following vehicle $i$ in the platoon, the source of the measurement noise is mainly from the movement state detection of the two predecessors $i-1$ and $i-2$. The measured position $X_{i-1}$ ($X_{i-2}$) of predecessor $i-1$ ($i-2$) consists of true value of position $\bar{X}_{i-1}$ ($\bar{X}_{i-2}$) and measurement noise $N_{i-1}$ ($N_{i-2}$): $X_{i-1} = \bar{X}_{i-1} + N_{i-1}$ ($X_{i-2} = \bar{X}_{i-2} + N_{i-2}$). From Fig. 5, the complementary sensitivity functions $T_{1,i}$ ($T_{2,i}$) can be used to describe the relationship between the processed position output $X_{pr,i}(s)$ of vehicle $i$ and measurement noise $N_{i-1}$ ($N_{i-2}$) included in position of predecessor $i-1$ ($i-2$).

$$T_{1,i}(s) = X_{pr,i}(s)N_{i-1}(s)^{-1} = \alpha_{b,i}(\zeta_i)H_i(s)G_i(s)K_i(s)[1 + H_i(s)G_i(s)K_i(s)]^{-1} \tag{37}$$

$$T_{2,i}(s) = X_{pr,i}(s)N_{i-2}(s)^{-1} = \beta_{b,i}(\zeta_i)H_i(s)G_i(s)K_i(s)[1 + H_i(s)G_i(s)K_i(s)]^{-1} \tag{38}$$

The magnitude of complementary sensitivity function $T_{1,i}$ ($T_{2,i}$) at a high frequency represents the effect of measurement noise mitigation (a larger value of $T_{1,i}$ or $T_{2,i}$ indicates reduced mitigation of measurement noise). The key aspect of mitigating high frequency measurement noise of $X_{i-1}$ ($X_{i-2}$) is to decrease the value of $T_{1,i}$ ($T_{2,i}$) in the high frequency band. Substituting $G_i(s)$, $H_i(s)$, and $K_i(s)$ from Equations (23), (28), and (33) into Equations (37)-(38), we have:

$$\lim_{s\to\infty} T_{1,i}(s) = \lim_{s\to\infty}\alpha_{b,i}(\zeta_i)\frac{h\omega_{K,i}s^2 + (h\omega_{K,i}^2 + \omega_{K,i})s + \omega_{K,i}^2}{(1 + h\omega_{K,i})s^2 + (h\omega_{K,i}^2 + \omega_{K,i})s + \omega_{K,i}^2} = \alpha_{b,i}(\zeta_i)\frac{h\omega_{K,i}}{1 + h\omega_{K,i}} \tag{39}$$

$$\lim_{s\to\infty} T_{2,i}(s) = \lim_{s\to\infty}\beta_{b,i}(\zeta_i)\frac{h\omega_{K,i}s^2 + (h\omega_{K,i}^2 + \omega_{K,i})s + \omega_{K,i}^2}{(1 + h\omega_{K,i})s^2 + (h\omega_{K,i}^2 + \omega_{K,i})s + \omega_{K,i}^2} = \beta_{b,i}(\zeta_i)\frac{h\omega_{K,i}}{1 + h\omega_{K,i}} \tag{40}$$

From Equations (18)-(21), we have $\alpha_{b,i}(\zeta_i) \leq 1$ and $\beta_{b,i}(\zeta_i) \leq 1$. By setting an upper bound for the product of $h\omega_{K,i}$ as: $h\omega_{K,i} \leq W_{max}$, the upper bounds of $\lim_{s\to\infty} T_{1,i}(s)$ and $\lim_{s\to\infty} T_{2,i}(s)$ can be determined as $\lim_{s\to\infty} T_{1,i}(s) \leq$ $h\omega_{K,i}/[1+h\omega_{K,i}] \leq W_{max}/[1+W_{max}]$ and $\lim_{s\to\infty} T_{2,i}(s) \leq h\omega_{K,i}/[1+h\omega_{K,i}] \leq W_{max}/[1+W_{max}]$, respectively, which indicates that the high frequency measurement noise from two predecessors is individually attenuated by at least $W_{max}/[1+W_{max}]$. This completes the proof for Lemma 1.

In terms of operational deployment, $W_{max}$ depends on the measurement noise mitigation needs of the specific system. In the study experiments, we aim to attenuate the high frequency measurement noise by a factor of at least 2/3. The upper bound of $h_d\omega_{K,i}$ is set as $W_{max} = 2$.

### 4.2.2. String stability analysis

The transfer function of string stability is specified as a measure of the signal amplification upstream from a platoon. The platoon head-to-tail string stability is analyzed in this study to ensure that the oscillations in downstream traffic are damped when they reach the tail of the platoon. The string stability transfer function is the ratio of the trajectory oscillation in the Laplace domain of the $i$th vehicle and the leading vehicle 0 in the platoon:

$$SS_{X,i}(s) = X_i(s)/X_0(s) \tag{41}$$

To ensure head-to-tail string stability, since $s = j\omega$, we have:

$$\left\|SS_{X,i}(j\omega)\right\|_\infty = \left\|X_i(j\omega)/X_0(j\omega)\right\|_\infty \leq 1 \tag{42}$$

where $j = \sqrt{-1}$, and the $\infty$-norm indicates that the magnitude of $\left|SS_{X,i}(j\omega)\right| \leq 1$ for all $\omega$. For simplification, we use $SS_{X,i}$, $X_i$, $G_i$, $H_i$, $K_i$, $F_{1,i}$, and $F_{2,i}$ to denote $SS_{X,i}(s)$, $X_i(s)$, $G_i(s)$, $H_i(s)$, $K_i(s)$, $F_{1,i}(s)$, and $F_{2,i}(s)$, respectively.

From Equations (1), (23), and (28)-(35), the transfer functions of string stability of all vehicles in platoon are described as:

$$\overline{SS} = T(\zeta)S \tag{43}$$

where $\overline{SS} = \left[SS_{x,0}\ SS_{x,1}\ SS_{x,2}\ SS_{x,3}\ \cdots\ SS_{x,n}\right]^T$, $S = [1\ 0\ 0\ 0\ \cdots\ 0]^T$, $SS_{x,0} = 1$ and $SS_{x,i} = X_i/X_0, i \geq 1$.

$$T(\zeta) = \begin{bmatrix} 1 & 0 & 0 & 0 & \cdots & 0 \\ \mathcal{G}_{1,1} & -1 & 0 & 0 & \ddots & \vdots \\ \mathcal{G}_{2,2} & \mathcal{G}_{2,1} & -1 & 0 & \ddots & 0 \\ 0 & \mathcal{G}_{3,2} & \mathcal{G}_{3,1} & -1 & \ddots & 0 \\ \vdots & \ddots & \ddots & \ddots & \ddots & 0 \\ 0 & \cdots & \cdots & \mathcal{G}_{n,2} & \mathcal{G}_{n,1} & -1 \end{bmatrix}^{-1}$$

$$\mathcal{G}_{i,1} = \alpha_{f,i}(\zeta_i)\Lambda_{f,i-1} + \alpha_{b,i}(\zeta_i)\Lambda_{b,i-1} \tag{44}$$

$$\mathcal{G}_{i,2} = \beta_{f,i}(\zeta_i)\Lambda_{f,i-2} + \beta_{b,i}(\zeta_i)\Lambda_{b,i-2} \tag{45}$$

where $\Lambda_{f,i-2} = \frac{G_iF_{2,i}s^2}{1+G_iK_iH_i}$ is the transfer function between the position of vehicles $i-2$ and $i$ with respect to the feedforward term $U_{f,i-2}(s)$; $\Lambda_{f,i-1} = \frac{G_iF_{1,i}s^2}{1+G_iK_iH_i}$ is the transfer function between vehicles $i-1$ and $i$ with respect to feedforward term $U_{f,i-1}(s)$; $\Lambda_{b,i-1} = \Lambda_{b,i-2} = \frac{G_iK_i}{1+G_iK_iH_i}$ is the transfer function between the position of vehicles $i-1$ or $i-2$ and $i$ with respect to the feedback term $U_{b,i}(s)$.

Based on the four communication statuses described in Fig. 4, we analyze the corresponding feasible regions for the time headway $h$ and cut-off frequency $\omega_{K,i}$ to ensure string stability, using Proposition 1 and Lemma 2.

**Proposition 1:** If $\sigma \geq 0$, a first order transfer function: $q(j\omega) = \frac{1}{1+j\omega\sigma}$ satisfies the string stability in Equation (42).

**Proof:** Since $|q(j\omega)| = \frac{1}{\sqrt{1+\sigma^2\omega^2}}$ and phase angle $\angle q(j\omega) = -\arctan(\sigma\omega)$, the magnitude of $q(j\omega)$ will always be lesser than or equal to one. However, when $\sigma < 0$, the phase angle of $q(j\omega)$ is positive, which indicates that the system corresponding to the first order transfer function is not practically deployable. Hence, $\sigma \geq 0$ is essential and necessary to guarantee string stability. This completes the proof for Proposition 1.

**Lemma 2**: The proposed adaptive PD controller in Section 4.1 can ensure the string stability of a platoon if the time headway $h$ and controller cut-off frequency $\omega_{K,i}$ satisfy the following conditions:

(i)   For the CACC cases, the time headway $h$ satisfies $h > 0$.

(ii)   For the ACC case, a lower bound for $h\omega_{K,i}$ satisfies $h\omega_{K,i} \geq \sqrt{2}$.

**Proof:** The string stability can be analyzed by performing an arithmetic operation on the transfer function of string stability, by checking if the ratio of trajectory oscillation in the frequency domain of the $i$th vehicle and the leading vehicle 0 is lesser than or equal to 1.

(1) The CACC cases:

From Equations (43)-(45), the transfer functions of string stability for the CACC cases and the ACC case are:

$$SS_{X,i} = \left(\alpha_{f,i}(\zeta_i)\Lambda_{f,i-1} + \alpha_{b,i}(\zeta_i)\Lambda_{b,i-1}\right)\frac{X_{i-1}}{X_0} + \left(\beta_{f,i}(\zeta_i)\Lambda_{f,i-2} + \beta_{b,i}(\zeta_i)\Lambda_{b,i-2}\right)\frac{X_{i-2}}{X_0} \quad (46)$$

Note that to directly analyze string stability transfer function is complex as it is a high-order transfer function. To address this problem, we consider only the worst case in Equation (46) where the values of both $X_{i-2}/X_0$ and $X_{i-1}/X_0$ are equal to one (implying head-to-tail marginally string stability, which means the traffic oscillation is neither amplified nor damped when it propagates in the traffic flow). This enables the determination of a more conservative, feasible region of the two parameters to ensure string stability. From Equations (18)-(20), when $X_{i-2}/X_0 = X_{i-1}/X_0 = 1$, Equation (46) becomes:

$$SS_{X,i} = \begin{cases} \left(\alpha G_i K_i + \beta G_i K_i + \alpha G_i F_{1,i}s^2 + \beta G_i F_{2,i}s^2\right)(1 + G_i K_i H_i)^{-1}, & CACC1 \\ \left(G_i K_i + G_i F_{1,i}s^2\right)(1 + G_i K_i H_i)^{-1}, & CACC2 \text{ and } CACC3 \end{cases} \quad (47)$$

Substituting for $H_i$ from Equation (30), Equation (47) can be simplified to:

$$SS_{X,i} = \begin{cases} [1 + (2 - \alpha)hs]^{-1}, & CACC1 \\ (1 + hs)^{-1}, & CACC2 \text{ and } CACC3 \end{cases} \quad (48)$$

To guarantee the string stability condition in Equation (42), the magnitude of $SS_{X,i}$ should not be greater than one. Correspondingly, for CACC1, according to proposition 1, the feasible region is: $h > 0$ since $1 \leq 2 - \alpha < 2$ and time headway is positive; for CACC2 and CACC3, similarly, the feasible region is: $h > 0$.

(2) The ACC Case:

Under the ACC schematic, we have $\alpha_{f,i}(\zeta_i) = \beta_{f,i}(\zeta_i) = \beta_{b,i}(\zeta_i) = 0$, and $\alpha_{b,i}(\zeta_i) = 1$ from Equation (21). The transfer function of string stability from Equation (46) will degrade to:

$$SS_{X,i} = \frac{X_i}{X_0} = \Lambda_{b,i-1}\frac{X_{i-1}}{X_0} = \frac{G_i K_i}{1 + G_i K_i H_i} = \frac{\omega_{K,i}s + \omega_{K,i}^2}{(1 + h\omega_{K,i})s^2 + \omega_{K,i}(1 + h\omega_{K,i})s + \omega_{K,i}^2} \quad (49)$$

Consequently, by substituting $s = j\omega$ in Equation (49), the string stability condition becomes:

$$|SS_{X,i}| = \left|\frac{j\omega_{K,i}\omega + \omega_{K,i}^2}{(\omega_{K,i} - (1 + h\omega_{K,i})\omega^2) + j\omega_{K,i}(1 + h\omega_{K,i})\omega}\right| \leq 1 \quad (50)$$

Solving inequality (50) leads to:

$$\omega_{K,i}^2\left(2 - h^2\omega_{K,i}^2\right)\left(1 + h\omega_{K,i}\right)^{-2} \leq \omega^2 \quad (51)$$

Since $\min_{\omega \geq 0}\omega^2 = 0$, the inequality (51) can be solved by letting $\omega^2 = 0$. Then, the string stability region of the controller cut-off frequency and headway time is:

$$h\omega_{K,i} \geq \sqrt{2} \quad (52)$$

**Remark 1:** Based on the above analysis, the decision-making process can be summarized as: (i) $h\omega_{K,i}$ has a specific upper-bound for mitigating measurement noise effects; $h\omega_{K,i} \leq W_{max} = 2$; and (ii) string stability requires $h > 0$ for the CACC1, CACC2 and CACC3 cases, and $h\omega_{K,i} \geq \sqrt{2}$ for the ACC case.

**Remark 2:** The additional parameter settings are: (i) the desired time headways $h$ in all controllers should be identical to preclude traffic oscillations that are generated through controller switching; and (ii) to simplify the problem, this study considers a homogeneous platoon, implying that all the vehicles have the same adaptive PD controller.

## 5. Algorithm to solve OPT-II

The section describes the algorithm to solve the IFT optimization model OPT-II, formulated in Section 4.2. The objective function in Equation (17) is complex because: (i) the decision variable $\boldsymbol{\xi}$ is a binary vector, which makes it a discrete integer optimization problem; (ii) the transfer function $SS_{X,i}^2(j\omega, \boldsymbol{\xi_d})$ depends on the controller used in the degeneration scenario $\boldsymbol{\xi_d}$, which implies different speed oscillation energies for the platoon under different degeneration scenarios; and (iii) the objective function requires the trajectory oscillations of the ambient traffic conditions in the frequency domain $X(j\omega)$ in the integrand to determine the expected speed oscillation energies. Together, these factors make it difficult to obtain a closed-form solution, entailing the need for a numerical solution.

To solve OPT-II numerically, the primary concern is computational complexity arising mainly due to platoon size.

For a platoon with $N+1$ vehicles, there are $2^{N+1}$ candidate IFTs. For an IFT with $b$ vehicles that activate "send" functionalities, there are $2^b$ degeneration scenarios. For each degeneration scenario, the computation of the control performance requires the sum of the speed oscillation energies in the frequency domain for each vehicle. Hence, the computational complexity can increase significantly as the platoon size increases, leading to the need for computation efficiency. Though the optimal IFT is not updated frequently, it should be determined as quickly as possible so that the platoon can adjust the IFT to the optimal one before ambient traffic conditions change significantly. To enhance computation efficiency, we first analyze the activation statuses of the "send" functionalities of the leading and the last vehicles in the platoon as a pre-processing step, which can preclude the consideration of a subset of candidate IFTs. Then, a two-step algorithm is proposed to efficiently search for the optimal IFT among the remaining candidate IFTs.

### 5.1 Activation status of "send" functionalities of the leading and last vehicles in the platoon

The "send" functionality of the communication device for the last (following) vehicle should always be deactivated as it only needs to receive information to maintain control performance. Deactivating the "send" functionality of this vehicle can decrease the probability of communication collision and improve the reliability of the V2V communication.

Next, we investigate the activation status of the "send" functionality of the leading vehicle. We first compare the control performances of CACCs 1, 2, 3 and ACC by Proposition 2, and then prove that the leading vehicle always needs to activate its "send" functionality, using Lemma 3 and Theorem 1.

From Equation (46), we note that the transfer function of string stability $SS_{X,i}$ of vehicle $i$ is a function of its communication status $\zeta_i$. In this subsection, we denote it more specifically as $SS_{X,i,\zeta_i}$ to reflect the transfer function of string stability under different controllers (CACCs 1, 2, 3, and ACC, for $\zeta_i \in \{1,2,3,4\}$, respectively).

**Proposition 2:** Based on the four possible communication statuses in Fig. 4, the magnitudes of transfer functions of string stability satisfy: $|SS_{X,i,1}| < |SS_{X,i,2}| < |SS_{X,i,4}|$, $|SS_{X,i,1}| < |SS_{X,i,3}| < |SS_{X,i,4}|$, indicating that CACCs can damp traffic oscillations better than ACC.

**Proof:** To compare the magnitudes of transfer function of string stability under the CACCs and the ACC, the cut-off frequency $\omega_{c,\zeta_i}$, $\zeta_i \in \{1,2,3,4\}$ of the transfer function is analyzed. According to Palm (2005), the cut-off frequency $\omega_{c,\zeta_i}$ is a corner frequency beyond which the logarithmic value of $X_i/X_0$ should be smaller than $-3.01$dB, which indicates that the energy of oscillation will be effectively damped. A smaller cut-off frequency $\omega_{c,\zeta_i}$ implies better string stability performance. The cut-off frequency $\omega_{c,i}$ of a transfer function $SS_{X,i,\zeta_i}(j\omega)$ can be calculated by solving Equation (53) as follows:

$$20\log|SS_{X,i,\zeta_i}(j\omega_{c,\zeta_i})| = -3.01\text{dB} \tag{53}$$

From Equations (47) and (48), the string stability transfer function for CACC1 is:

$$SS_{X,i,1}(j\omega_{c,1}) = (\alpha+\beta)(1+G_iK_iH_i)[H_i(1+G_iK_iH_i)]^{-1} = H_i^{-1} = [1+j(2-\alpha)h\omega_{c,1}]^{-1} \tag{54}$$

By substituting Equation (54) into Equation (53) and solving it, we obtain the corresponding cut-off frequency in Equation (55), where $C = 10^{\frac{-3.01}{10}}$:

$$\omega_{c,1} = \sqrt{(1-C)[(2-\alpha)^2Ch^2]^{-1}} \tag{55}$$

From Equations (47) and (48), the transfer functions for CACC2 and CACC3 are:

$$SS_{X,i,2}(j\omega_{c,2}) = SS_{X,i,3}(j\omega_{c,3}) = (G_iK_i + G_iF_{1,i}s^2)(1+G_iK_iH_i)^{-1} = H_i^{-1} = (1+j\omega_{c,2}h)^{-1} \tag{56}$$

By substituting Equation (56) into Equation (53) and solving it, we obtain the corresponding cut-off frequency:

$$\omega_{c,2} = \omega_{c,3} = \sqrt{(1-C)[Ch^2]^{-1}} \tag{57}$$

From Equation (18), we have $0 < \alpha < 1$. By comparing Equations (55) and (57), we obtain that $\omega_{c,1} < \omega_{c,2} = \omega_{c,3}$. Then, we only need to compare $\omega_{c,2}$ or $\omega_{c,3}$ with $\omega_{c,4}$. From Equation (49), the transfer function of ACC is:

$$SS_{X,i,4}(j\omega_{c,4}) = \frac{G_iK_i}{1+G_iK_iH_i} = \frac{j\omega_{K,i}\omega_{c,4} + \omega_{K,i}^2}{j(\omega_{K,i}^2 + h\omega_{K,i})\omega_{c,4} + \omega_{K,i}^2 - (1+h\omega_{K,i})\omega_{c,4}^2} \tag{58}$$

By substituting Equation (58) into Equation (53) and solving it, we obtain the corresponding cut-off frequency:

$$\omega_{c,4} = \sqrt{\left[B + \sqrt{B^2 - 4C(1+h\omega_{K,i})^2(\omega_{K,i}^4(C-1))}\right]\left[2C(1+h\omega_{K,i})^2\right]^{-1}} \tag{59}$$

where $B = C(\omega_{K,i} + h\omega_{K,i}^2)^2 - 2C\omega_{K,i}^2(1+h\omega_{K,i}) - \omega_{K,i}^2$.

The inequality $\omega_{c,4} > \omega_{c,2}$ yields a lower bound for $h\omega_{K,i}$: $h\omega_{K,i} > (1+\sqrt{3})/2$. By considering the lower and upper bounds from Remark 1, $h\omega_{K,i}$ lies within $\left[\sqrt{2}, 2\right]$, which satisfies $\omega_{c,4} > \omega_{c,2}$.

We note that the cut-off frequencies of string stability under the CACC cases and the ACC case satisfy $\omega_{c,1} < \omega_{c,2} = \omega_{c,3} < \omega_{c,4}$. Hence, the magnitudes of the string stability transfer functions satisfy: $|SS_{X,i,1}| < |SS_{X,i,2}| < |SS_{X,i,4}|$, $|SS_{X,i,1}| < |SS_{X,i,3}| < |SS_{X,i,4}|$.

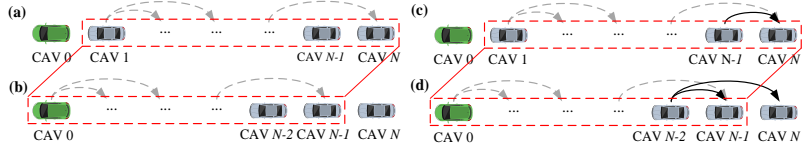

Fig. 6. Illustration of the deactivated or activated status of the "send" functionality of CAV 0.

Next, the "send" status of the leading vehicle is analyzed. Two IFTs are illustrated in Fig. 6(a) and 6(c). In both scenarios, the "send" functionality of the leading vehicle and the last vehicle are deactivated. In Fig. 6(a), vehicle $N-1$ has its communication device deactivated, while in Fig. 6(c) vehicle $N-1$ broadcasts messages. In Fig. 6(b) (Fig. 6(d)), the "send" activation status of vehicle $i$ (for $i = 0, 1, \cdots, N-1$) is the same as for vehicle $i+1$ in Fig. 6(a) (Fig. 6(c)). The "send" functionality is deactivated for the last vehicles in Fig. 6(b) and Fig. 6(d).

**Lemma 3:** In Fig. 6, the expected platoon speed oscillation energy of the IFT in Fig. 6(b) is always lower than that in Fig. 6(a). Similarly, the speed oscillation energy of the IFT in Fig. 6(d) is lower than that in Fig. 6(c).

**Proof:** Denote the IFTs in Figs. 6(a) and 6(b) as $\boldsymbol{\xi}^{(a)}$ and $\boldsymbol{\xi}^{(b)}$, and degeneration scenarios of $\boldsymbol{\xi}^{(a)}$, $\boldsymbol{\xi}^{(b)}$ as $\boldsymbol{\xi}_d^{(a)}$, $\boldsymbol{\xi}_d^{(b)}$. For each $\boldsymbol{\xi}_d^{(a)}$, there is a $\boldsymbol{\xi}_d^{(b)}$ whose IFT from vehicle 0 to vehicle $N-1$ is the same as the IFT of $\boldsymbol{\xi}_d^{(a)}$ from vehicle 1 to vehicle $N$. Hence, the probability of $\boldsymbol{\xi}_d^{(a)}$, $P_d(\boldsymbol{\xi}_d^{(a)})$, is equal to the probability of $\boldsymbol{\xi}_d^{(b)}$, $P_d(\boldsymbol{\xi}_d^{(b)})$ (i.e. $P_d(\boldsymbol{\xi}_d^{(a)}) = P_d(\boldsymbol{\xi}_d^{(b)})$). Then, we only need to compare the control performance of $\boldsymbol{\xi}_d^{(a)}$ and $\boldsymbol{\xi}_d^{(b)}$. Note that vehicle 1 in Fig. 6(a) and vehicle $N$ in Fig. 6(b) operate under ACC. To simplify notation, we use $SS_{X,i,\zeta_i}(\boldsymbol{\xi}_d^{(a)})$ and $X_0$ to denote $SS_{X,i,\zeta_i}(j\omega, \boldsymbol{\xi}_d^{(a)})$ and $X_0(j\omega)$, respectively. We denote $SS'_{X,i,\zeta_i}(\boldsymbol{\xi}_d^{(a)}) = X_i/X_1$; then, according to Equations (11) and (41), we have:

$$
\begin{aligned}
E_d(\boldsymbol{\xi}_d^{(a)}) &= 4\pi^2 \int_0^{+\infty} \left( \sum_{i=0}^{N} SS_{X,i,\zeta_i}^2(\boldsymbol{\xi}_d^{(a)}) \right) \omega^2 X_0^2 \, d\omega \\
&= 4\pi^2 \int_0^{+\infty} \left( 1 + SS_{X,1,4}^2 \sum_{i=1}^{N} SS'^2_{X,i,\zeta_i}(\boldsymbol{\xi}_d^{(a)}) \right) \omega^2 X_0^2 \, d\omega
\end{aligned}
\tag{60}
$$

$$
E_d(\boldsymbol{\xi}_d^{(b)}) = 4\pi^2 \sum_{i=0}^{N} \int_0^{+\infty} \omega^2 SS_{X,i,\zeta_i}^2(\boldsymbol{\xi}_d^{(b)}) X_0^2 \, d\omega = 4\pi^2 \int_0^{+\infty} \left( \sum_{i=0}^{N-1} SS_{X,i,\zeta_i}^2(\boldsymbol{\xi}_d^{(b)}) + SS_{X,n,4}^2(\boldsymbol{\xi}_d^{(b)}) \right) \omega^2 X_0^2 \, d\omega
\tag{61}
$$

$$
\frac{E_d(\boldsymbol{\xi}_d^{(b)}) - E_d(\boldsymbol{\xi}_d^{(a)})}{4\pi^2} = \int_0^{+\infty} \left( \sum_{i=0}^{N-1} SS_{X,i,\zeta_i}^2(\boldsymbol{\xi}_d^{(b)}) + SS_{X,n,4}^2(\boldsymbol{\xi}_d^{(b)}) - 1 - SS_{X,1,4}^2 \sum_{i=1}^{N} SS'^2_{X,i,\zeta_i}(\boldsymbol{\xi}_d^{(a)}) \right) \omega^2 X_0^2 \, d\omega
\tag{62}
$$

Since all vehicles have the same adaptive PD controller (Remark 2), we have $SS_{X,1,4}^2 = \Lambda_{b,0}^2 = \Lambda_{b,i-1}^2$ for any $i = 2, \dots, N$, where $\Lambda_{b,i-1}$ is as shown in Equation (49). The receiver state $\zeta_i$ of vehicle $i$ (for $i = 1, 2, \cdots, N$) in Fig. 6(a) is equal to the receiver state $\zeta_{i-1}$ of vehicle $i-1$ in Fig. 6(b), which indicates that $SS'^2_{X,i,\zeta_i}(\boldsymbol{\xi}_d^{(a)}) = SS_{X,i-1,\zeta_{i-1}}^2(\boldsymbol{\xi}_d^{(b)})$. From Proposition 2 and Equations (41) and (49), for $i > 3$, we have $SS_{X,i-1,\zeta_{i-1}}^2(\boldsymbol{\xi}_d^{(b)}) \leq SS_{X,i-2,\zeta_{i-2}}^2(\boldsymbol{\xi}_d^{(b)}) \Lambda_{b,i-2}^2 \leq SS_{X,i-3,\zeta_{i-3}}^2(\boldsymbol{\xi}_d^{(b)}) \Lambda_{b,i-2}^2 \Lambda_{b,i-3}^2 \leq \cdots < SS_{X,1,\zeta_1}^2(\boldsymbol{\xi}_d^{(b)}) \prod_{i'=2}^{i} \Lambda_{b,i'-1}^2 < \prod_{i'=1}^{i} \Lambda_{b,i'-1}^2 = \Lambda_{b,0}^{2i}$. For $i = 1, 2$, we have $SS_{X,i-1,\zeta_i}^2(\boldsymbol{\xi}_d^{(b)}) < \prod_{i'=1}^{i} \Lambda_{b,i'-1}^2 = \Lambda_{b,0}^{2i}$. These inequalities indicate that the string stability performance of vehicle $i$ under degeneration scenario $\boldsymbol{\xi}_d^{(b)}$ is always better than the performance when all its predecessors deactivate their "send" functionalities. Accordingly, we have:

$$
\sum_{i=0}^{N-1} SS_{X,i,\zeta_i}^2(\boldsymbol{\xi}_d^{(b)}) + SS_{X,n,4}^2(\boldsymbol{\xi}_d^{(b)}) - 1 - SS_{X,1,4}^2 \sum_{i=1}^{N} SS'^2_{X,i,\zeta_i}(\boldsymbol{\xi}_d^{(a)})
\tag{63}
$$

$$< \left(1 - \Lambda_{b,0}^2\right)\left(\sum_{i=1}^{N-1} \Lambda_{b,0}^{2i}\right) + \Lambda_{b,0}^{2N} - 1 \ = \left(1 - \Lambda_{b,0}^2\right)\frac{1 - \Lambda_{b,0}^{2N}}{1 - \Lambda_{b,0}^2} + \Lambda_{b,0}^{2N} - 1 = 0$$

By substituting inequality (63) into Equation (62), we have $E_d\left(\xi_d^{(b)}\right) < E_d\left(\xi_d^{(a)}\right)$, which indicates that each degeneration scenario $\xi_d^{(b)}$ of IFT $\xi^{(b)}$ has lower oscillation energy than the corresponding degeneration scenario $\xi_d^{(a)}$ of IFT $\xi^{(a)}$. Since the corresponding probabilities of each degeneration scenario pair are identical, $\xi^{(b)}$ always outperforms $\xi^{(a)}$. Similarly, we can prove that the IFT in Fig. 6(d) outperforms that in Fig. 6(c).

**Theorem 1**: In the optimal IFT, the "send" functionality of the leading vehicle is always activated.

**Proof:** If the "send" functionality of leading vehicle is not activated, then from Lemma 3 we can always change the IFT from (a) to (b) or from (c) to (d) to find a better one until the leading vehicle activates the "send" functionality.

### 5.2 Two- step algorithm for solving OPT-II

Different IFTs may have several identical degeneration scenarios. For example, IFT $[0, 0, 1, 0, 0]$ in Fig. 2(a) is a degeneration scenario of IFT $[1, 0, 1, 0, 0]$ illustrated in Fig. 2(a). However, it is also a degeneration scenario of IFT $[1, 1, 1, 0, 0]$ when both vehicles 0 and 1 fail to send messages at the same time. Since the leading vehicle always needs to activate its "send" functionality and the last vehicle needs to deactivate it, a fully-activated IFT except for the last vehicle ($\xi = [1, \dots, 1,0]$) includes all possible degeneration scenarios for other IFTs. Therefore, we only need to investigate the string stability performance of the degeneration scenarios for that IFT. For the other IFTs, we will just use the string stability performance of the relevant degeneration scenarios calculated for the fully-activated IFT.

Motivated by above observation, we propose a two-step algorithm. The first step calculates the string stability performance of degeneration scenarios for the fully-activated IFT to construct a control performance table according to Sections 3.1 and 4. Then, in the second step, for each $\xi \in \Omega$, we traverse all possible degeneration scenarios $\xi_d \in \Omega_d(\xi)$, and add the corresponding control performances from the table generated in the first step with a weight $P_d\left(\xi_d(\xi)\right)$ formulated from the contention model in Section 3.2 to obtain the expected string stability of IFT $\xi$. The pseudo code of the two-step algorithm is shown in the following table:

| Step 1 |
| --- |
| **input** ambient traffic oscillation in the frequency domain $X(j\omega)$, average density $\bar{k}$, and platoon size $N + 1$; |
| **set** $\xi = [1, \dots, 1,0] \in \mathbb{R}^{N+1}$, update $\Omega_d(\xi)$ |
| **for** any $\xi_d(\xi) \in \Omega_d(\xi)$ |
|     Determine its speed oscillation energies $E_d\left(\xi_d(\xi)\right)$ using Equation (11). |
| **end** |
| **output** $\hat{\xi}_d = \xi_d(\xi)$ and corresponding speed oscillation energies $\hat{E}_d\left(\hat{\xi}_d\right) = E_d\left(\xi_d(\xi)\right)$ for any $\xi_d(\xi) \in \Omega_d(\xi)$ |
| **Step 2** |
| **input** $E(x)$ |
| **set** the set of candidate IFTs $\Omega = \{[1, \eta_1, \dots, \eta_{N-1}, 0] | \eta_i \in \{0, 1\}$ for $i = 1, \dots, N-1\}$ |
| **initialize** the optimal expected speed oscillation energies $E^* \leftarrow +\infty$ |
| **for** any $\xi \in \Omega$ |
|     Update $\Omega_d(\xi)$ |
|     **for** any $\xi_d(\xi) \in \Omega_d(\xi)$ |
|         Determine the $P_d\left(\xi_d(\xi)\right)$ using Equation (16). |
|         Find $\hat{\xi}_d = \xi_d(\xi)$, then $E_d\left(\xi_d(\xi)\right) \leftarrow \hat{E}_d\left(\hat{\xi}_d\right)$. |
|     **end** |
|     Determine the expected speed oscillation energies $E(\xi)$ under the IFT $\xi$ using Equation (2). |
|     **if** $E(\xi) < E^*$ |
|         Update $E^* \leftarrow E(\xi)$ and the optimal IFT $\xi^* \leftarrow \xi$ |
|     **end** |
| **end** |
| **output** $\xi^*$ |

## 6. Numerical experiments

### 6.1 Experiment design and parameter setting

Numerical experiments are conducted to analyze the CACC-OIFT strategy. First, the performance of V2V communication and the computational efficiency of the algorithm are investigated. Next, the performance of CACC-OIFT is analyzed. The optimization procedure and the performance comparison simulations are conducted on a C++ platform that integrates network simulator NS-3 to emulate the V2V communication process.

The experiment setup consists of a $N+1$ CAV platoon with one leading vehicle ($i=0$) and $N$ following vehicles ($i=1,...,N$, and $N=11,...,15$). The movement of the leading vehicle is predetermined according to NGSIM field data (US DOT, 2007), which contains a 240-second vehicle trajectory on eastbound I-80 in the San Francisco Bay area at Emeryville, California. The frequency domain trajectory oscillation $X(j\omega)$ and average density of ambient traffic flow $\bar{k}$ are provided to the optimization model. The first following vehicle ($i=1$) receives information only from one preceding vehicle ($i=0$); so, the controller will switch between CACC2 and ACC if the IFT degenerates. For the other vehicles ($i=2,...,N$), the controller can switch among the four controllers (i.e. CACC 1, 2, 3, and ACC). The desired time headways in all controllers are set to $h=1s$. The cut-off frequency $\omega_{K,i}$ is set as 0.8, 0.8, 0.9 and 1.45 for CACC 1, 2, 3, and ACC, respectively. The parameters $\alpha$ and $\beta$ are set as 0.7 and 0.3. The control time interval is set as 0.1s. The network parameters are listed in Table 1.

Table 1. Network parameters.

| Parameter | Value | Parameter | Value | Parameter | Value |
|---|---|---|---|---|---|
| Communication range | 0.2 km | Information generation rate | 10 Hz | Data rate | 3 Mbps |
| Packet size | 500 B | Contention window size | 8 | Slot time | 16 μs |

The performance of CACC-OIFT is analyzed by comparing it with two other control strategies. The three strategies are: (i) CACC-OIFT, which includes the IFT optimization from Section 3 and the adaptive PD controller from Section 4; (ii) CACC-DIFT, which includes the adaptive PD controller from Section 4 with a fully-activated IFT; (iii) CACC with a fixed IFT (CACC-FIFT), which includes the CACC and ACC schemes developed in Naus et.al. (2010). We also analyzed another CACC controller with a fixed IFT (Schakel et.al., 2010). However, as its string stability is not guaranteed, it performs worse than the CACC in Naus et.al. (2010). Hence, hereafter, we focus on the CACC-FIFT developed by Naus et.al. (2010).

### 6.2 Performance of V2V communications, optimization result and computation efficiency

To illustrate the performance of V2V communications under different ambient traffic conditions, we conduct simulations with different parameters. Fig. 7 shows the communication success rates under different ambient traffic conditions and different IFTs from the simulations in NS-3. In Fig. 7(a), the x-axis denotes the average traffic density $\bar{k}$. This study sets the range of average density from 25 vehicles/km to 40 vehicles/km. The success rate decreases with the increase in the percentage of vehicles with activated 'send' functionalities of V2V communication devices in communication range. A higher percentage of activated "send" functionalities leads to more intense contentions for the chance to broadcast. For the same proportion of activated "send" functionalities in communication range, a higher traffic density will result in a higher failure probability. This is because there are more vehicles in the communication range of each vehicle if the average density of ambient traffic flow is higher. Hence, there are more vehicles with activated "send" functionalities when the proportions are identical. Fig. 7(b) compares the simulation results under $\bar{k}=25$ with the contention model (discussed in Section 3.2) that is calibrated for the IFT optimization model. The mean error is -0.25% and the standard deviation of the error is 0.0526. Similar results are observed under different average densities, implying that the model can accurately describe the success rate under different percentages of activated "send" functionalities. We do not show all results due to the page limit.

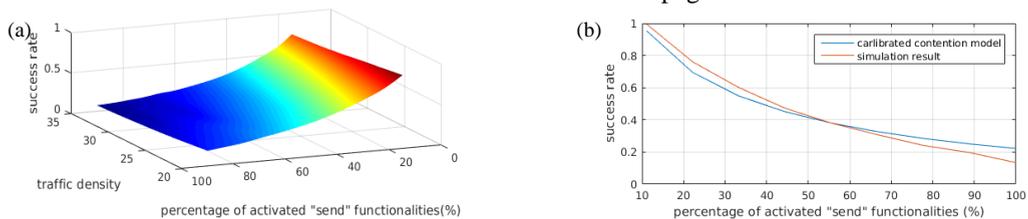

Fig. 7. Experiment results for V2V communications: (a) Communication success rates under different ambient traffic conditions; (b) Comparison of calibrated contention model and simulation results for *k*=20.

Table 2 illustrates the optimal IFTs under different ambient traffic conditions and platoon sizes. It can be observed that the optimal IFTs are in accordance with Theorem 1 and Corollary 1. Consecutive vehicles with activated "send' functionalities can efficiently increase the occurrence of CACC cases, especially the CACC1 case which has much better control performance than the ACC case. However, the success rate of communication decreases since it increases the probability of information collision. Thereby, there are some consecutive vehicles with deactivated communication devices directly following those with activated ones, such as vehicles 3, 4, and 5 for the scenario with $\bar{k}$=25 and N=14. For a given platoon size, the number of consecutive vehicles with deactivated "send' functionalities is bigger if the density $\bar{k}$ is higher. For a given $\bar{k}$, there exist several activated/deactivated patterns of communication devices in the platoon. For example, the same pattern 111000 exists for the first 6 and last 6 vehicles in the scenario $\bar{k}$=25, and N=14, as the communication environment and control scheme are similar for a pattern.

Table 2. Network parameters.

| N=14 | Optimal IFT | $\bar{k}$=25 | Optimal IFT |
|---|---|---|---|
| $\bar{k}$=25 | 111000111000110 | N=11 | 111000111000 |
| $\bar{k}$=30 | 111000001110000 | N=12 | 111000111000 |
| $\bar{k}$=35 | 110000001110000 | N=13 | 111000111110000 |
| $\bar{k}$=40 | 110000000110000 | N=14 | 111000111000110 |
| | | N=15 | 111000111000110 |

To enable the practical deployment of CACC-OIFT, computational efficiency should be verified. Though we improve the algorithmic efficiency in several parts as discussed in Section 5, it is still a method of exhaustion; the computational time increases with platoon size. Van Arem et al. (2006) suggest that the length of a platoon could be limited under traffic merge conditions. In our numerical experiments, the optimization procedure for a platoon with 15 vehicles only takes 48.23 seconds on a PC with Intel E3-1505M 2.80GHz 8Gb. Further, in practice, parallel computing can be enabled by vehicle due to the problem characteristics. Then, computational time for this platoon reduces to 3.22 seconds. Since the optimal IFT is updated every time period (e.g., 5 minutes) or when ambient traffic oscillation conditions change significantly, this small computation time ensures the practicality of CACC-OIFT.

*6.3 Control performance evaluation*

Here, we compare the performance of CACC-OIFT with those of CACC-DIFT and CACC-FIFT. The experiments analyze the three controllers in the context of unreliable V2V communications by simulating in NS-3. A 15-CAV platoon is analyzed in a traffic flow with average density 28.57 vehicle/km for 240s. The ambient traffic conditions do not change significantly. Under CACC-OIFT, the vehicle platoon will follow the IFT from the optimization model (111000111000110). Fig. 8 shows the spacing speed tracking error between adjacent vehicles in the platoon under these controllers, and Fig. 9 shows the standard deviations of the spacing and speed tracking errors, and vehicle speed.

Fig. 8 illustrates that the spacing tracking error of vehicles is mitigated based on their positions in the platoon. The figure shows that CACC-OIFT outperforms the other two controllers. For example, the maximum spacing tracking error of the second following vehicle ($i$=2) under CACC-OIFT is 1.05m, compared with 1.42m for CACC-DIFT and 1.51m for CACC-FIFT. For the last following vehicle ($i$=14), the maximum spacing tracking errors are 0.37m, 0.68m and 0.79m for CACC-OIFT, CACC-DIFT and CACC-FIFT, respectively. The standard deviations of spacing and speed tracking errors are compared for the three controllers in Fig. 9(a) and Fig. 9(b), respectively. Fig. 9(a) shows that the standard deviation of spacing tracking error decreases sequentially across vehicles in the platoon for all controllers. However, CACC-OIFT performs better than the other two controllers as its spacing error reduces the quickest. Further, the profile of the spacing error standard deviation cycles from steep to flat. For example, the spacing error is reduced significantly for first 4 vehicles and then is almost constant for vehicles $i$=5 and $i$=6. This is because the IFT optimization deactivates the "send' functionalities of V2V communication devices for several vehicles. Thus, some vehicles will operate under the ACC case, which does not leverage V2V connectivity. However, these deactivations lead to more reliable V2V communication connections for remaining links in CACC-OIFT and the consequent significant tracking error reduction for the vehicles. A similar trend is observed in Fig. 9(b) which shows the standard deviation of the speed tracking error.

To further investigate the performance benefits under CACC-OIFT, the performance of the three CACC control strategies is compared when traffic oscillates (e.g., stop-and-go or slow-and-fast traffic). The standard deviations of the vehicle speed are shown in Fig. 9(c). It illustrates that the fluctuation in standard deviation of speed decreases under all three schemes as the tail of the platoon is approached, which implies that traffic oscillations are damped. Further, CACC-OIFT reduces the speed fluctuations significantly as it proactively leverages the dynamic nature of the IFT. In summary, we conclude that the performance of a CAV platoon controlled by the proposed CACC-OIFT is better and more robust than that of the other two controllers in a realistic V2V communications environment. Also, based on the discussion in Section 1, CACC-DIFT performs better than CACC-FIFT as it considers IFT dynamics, albeit passively, unlike CACC-FIFT which assumes a fixed IFT.

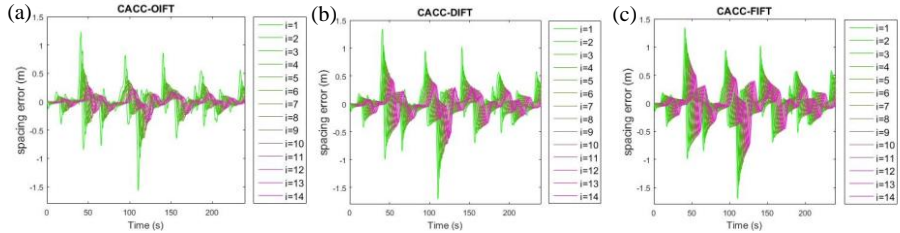

Fig. 8. Spacing tracking error under different controllers: (a) CACC-OIFT; (b) CACC-DIFT; (c) CACC-FIFT.

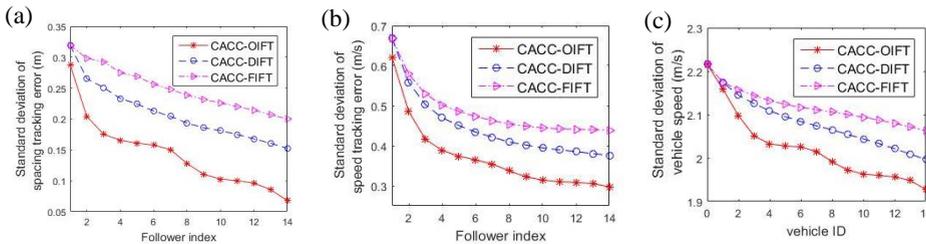

Fig. 9. Standard deviation of: (a) spacing tracking error; (b) speed tracking error; (c) vehicle speed.

## 7. Concluding comments

This study proposes a novel CACC strategy, CACC-OIFT, to explicitly factor IFT dynamics and to leverage it to enhance the platoon performance in an unreliable V2V communication context for a pure CAV platoon. The proposed CACC-OIFT consists of an adaptive PD controller and an IFT optimization model. Given the adaptive PD controller with a two-predecessor-following scheme, and the ambient traffic conditions and the platoon size just before the start of a time period, the IFT optimization model determines the optimal IFT that dynamically activates and deactivates the "send" functionality of the V2V communication devices of all vehicles in platoon, which maximizes the expected string stability. Since communication failures can cause IFT to degenerate dynamically, all possible degeneration scenarios for that IFT are considered in this expectation. The degeneration scenario probabilities are determined based on the communication failure probabilities for that time period which depend on the ambient traffic conditions. In the operational deployment context, based on the various degeneration scenarios for the optimal IFT at different time instants within the time period, the adaptive PD controller continuously determines the car-following behaviors of the vehicles in the platoon. A two-step algorithm is proposed to solve the IFT optimization problem by leveraging some key proven properties, such as the leading vehicle in the platoon should always activate its "send" functionality. Extensive numerical simulations are conducted in NS-3 to illustrate the effectiveness of CACC-OIFT.

To the best of our knowledge, this is the first attempt to explicitly factor IFT dynamics and to leverage it to enhance the performance of CACC strategies. Further, it is the first study to perform a rigorous mathematical modeling of the problem to theoretically illustrate properties. The insights from numerical experiments suggest that CACC-OIFT can leverage IFT dynamics to proactively reduce V2V communication failures while ensuring realism in terms of factoring the ambient traffic conditions. Further, the proposed two-step algorithm and its ability to be parallelized ensure computational tractability for operational deployment for platoons of considerable size (15 vehicles in this study). Also, the study insights provide key pointers for future CACC designs, in that communication failures and IFT dynamics should be considered to enable realism and enhance control performance.

In summary, CACC-OIFT can generate a more reliable IFT for a CAV platoon, damp traffic oscillation propagation, and stabilize the traffic flow more efficiently for the entire platoon. Thereby, CACC-OIFT is string stable and outperforms strategies proposed in the current literature, CACC-DIFT and CACC-FIFT, considerably.

The proposed modeling approach also provides opportunities for improvements beyond the current study, including: (i) considering the role of receiver failure in V2V communications; (ii) considering actuator delay, nonlinear vehicle dynamics and external disturbances in controller design; (iii) implementing the IFT optimization model in other CACC strategies with potential for better string stability performance (such as sliding model control and model predictive control); and (iv) designing a CACC control specifically for a mixed flow environment, since CACCs in the existing mixed traffic flow literature assume a pure CAV platoon, though a CACC can be used in mixed flow context.